\documentclass[aps,prd,amssymb,showpacs,twocolumn,superscriptaddress,nofootinbib,letterpaper]{revtex4}
\usepackage{setspace}
\usepackage{graphicx}
\usepackage{psfrag}
\usepackage{amsmath}
\usepackage{float}
\def\DERIV#1#2{\frac{d{#1}}{d{#2}}}
\def\PDERIV#1#2{\frac{\partial{#1}}{\partial{#2}}}
\def\grad{\nabla}
\def\alphadot{\frac{\partial\alpha}{\partial t}}
\def\adot{\frac{\partial a}{\partial t}}
\def\aprime{\frac{\partial a}{\partial r}}
\def\alphaprime{\frac{\partial\alpha}{\partial r}}

\def\bi{\begin{itemize}}
\def\ei{\end{itemize}}
\def\be{\begin{equation}}
\def\ee{\end{equation}}
\def\bean{\begin{eqnarray}}
\def\eean{\end{eqnarray}}

\begin{document}
\title{Critical collapse in the spherically-symmetric Einstein-Vlasov model}

\author{Arman Akbarian}
\affiliation{Department of Physics and Astronomy,
     University of British Columbia,
     Vancouver BC, V6T 1Z1 Canada}

\author{Matthew W. Choptuik}
\affiliation{CIFAR Cosmology and Gravity Program \\
     Department of Physics and Astronomy,
     University of British Columbia,
     Vancouver BC, V6T 1Z1 Canada}

\begin{abstract}
We solve the coupled Einstein-Vlasov system in spherical symmetry using direct
numerical integration of the Vlasov equation in phase space.
Focusing on the case of massless particles we study critical 
phenomena in the model,
finding strong evidence for generic type I behaviour at the black hole
threshold that parallels what has previously been observed in the massive
sector.  For differing families of initial data we find 
distinct critical solutions, so there is no universality of the critical 
configuration itself.  However we find indications of 
at least a weak universality in the lifetime 
scaling exponent, which is yet to be understood.  
Additionally, we clarify the role that angular momentum plays in 
the critical behaviour in the massless case.
\end{abstract}


\maketitle

\section{Introduction\label{intro}}

In this paper we report results from an investigation of critical collapse in the spherically 
symmetric Einstein-Vlasov system, which describes the interaction of collisionless
matter with the general relativistic gravitational field.  After more than two decades of study,
the field of black hole critical phenomena has matured and although we present a brief overview 
below, we assume that the reader is 
at least somewhat familiar with the key concepts and results in the subject: those who are not
can consult comprehensive review articles~\cite{Gundlach:1997wm,Gundlach:2007lrr}.

We recall that
critical phenomena can be identified in a given model by considering dynamical
evolution of initial data that is characterized by a parameter, $p$, such that 
for sufficiently small $p$ the gravitational interaction remains weak and the matter (or gravitational
energy in the vacuum case) typically disperses, while for sufficiently large $p$ a black hole
forms.  By tuning $p$ between these limits we isolate a critical parameter value $p^\star$ that
generates a solution representing the threshold of black hole formation for the particular 
family of initial data.   The behaviour that arises 
in the near-critical regime $p\to p^\star$ constitutes what is meant by 
black hole critical phenomena.  Depending on the particulars of the model, these 
phenomena will comprise one or more of the following: 1) existence of a special
solution at criticality with possible universality with respect to the parameterization
of the initial data, 2) symmetry of the critical solution beyond any imposed in the 
model itself and 3) scaling of dimensionful physical quantities as a function of $|p-p^\star|$,
with scaling exponents which may also be universal in the sense given above. 
These properties can largely be explained by observing that a critical solution
has a single unstable mode in perturbation theory, whose associated eigenvalue (Lyapunov 
exponent) can be immediately related to the empirically measured scaling exponent.

For the most part, the critical transitions that have been observed to date 
fall into two classes that are dubbed type I and type II in analogy
with first and second order phase transitions, respectively, in statistical mechanical systems, and 
where the behaviour of the black hole mass plays the role of an order parameter.
A type I transition is characterized by a static or periodic critical
solution, with a scaling law 
\begin{equation}
\tau = -\sigma\ln|p-p^\star| \, .
\label{eq:deftimescaling}
\end{equation}
Here, $\tau$ is the lifetime of the near-critical configuration---the amount of time 
that the dynamical configuration is closely approximated by the precisely critical 
solution---and the scaling exponent, $\sigma$, is the reciprocal of the Lyapunov exponent, $\lambda$,
associated with the solution's single unstable mode.
In this case the black hole mass is {\em finite} at threshold since when the marginally 
stable static or periodic solution collapses, most of its mass-energy will end up inside
the horizon.

Previous studies~\cite{Rein-Randall:1998,Inaki:2001,stevenson,And_Rein:2006} 
have strongly suggested that the critical behaviour in the Einstein-Vlasov
model is generically type I and  our current results bear this out.  So far as we know, type II collapse,
where the critical solution is self similar and the black hole mass is infinitesimal 
at threshold, is not relevant to the model and will not be considered here.


In the Einstein-Vlasov system the matter
content of spacetime is specified by a density function $f(t,x^i,p_j)$ in phase space 
whose evolution is given by the Vlasov equation, while the geometry is governed by the 
Einstein equations.  
Numerical studies of the model have a long history, dating back to the 
work by Shapiro and Teukolsky, both in spherical symmetry~\cite{shapiro:1985-1,shapiro:1985-2,shapiro:1985-3} and 
axisymmetry~\cite{Shapiro:1991zza,Shapiro92a}.  Investigation of critical collapse in the spherically symmetric 
sector was initiated by Rein et al~\cite{Rein-Randall:1998}
who observed finite black hole masses at threshold for all families considered.
Subsequent work by Olabarrieta and Choptuik~\cite{Inaki:2001} corroborated these findings and
additionally provided evidence that the threshold solutions were static with lifetime 
scaling of the form~(\ref{eq:deftimescaling}). Moreover, there were some indications in this 
latter study that there might 
be a universal critical solution and associated scaling exponent.  

More recently, Andr\'{e}asson and Rein have carried out a comprehensive study of 
precisely static solutions of the model, concentrating on their stability both 
generally and in the context of critical phenomena~\cite{And_Rein:2006,And_Rein:2007}.
Many of their observations and results are pertinent to 
our current investigation.
First, they point out that static solutions can be constructed via 
a specific ansatz for the distribution function that is discussed in Sec.~\ref{sec:initstat}.
Second, using this ansatz they construct parameterized sequences of static solutions, 
and, following astrophysical practice, characterize the solutions 
by their central redshifts and binding energies.  Third, they present strong 
evidence that a maximum in the binding energy along a sequence signals an onset of instability and 
that at least some of the configurations that lie along an unstable branch can act 
as type I solutions in the critical collapse context.  This immediately 
establishes that there can {\em not} be universality in the model.  Fourth, and finally, 
they show that dispersal is not the only stable end state of sub-critical collapse, but that 
relaxation to a bound state is also possible, contingent on the sign of the binding energy.
Overall, the picture of critical behaviour that emerges very much parallels that which is 
observed for type I transitions in the perfect-fluid and massive-scalar 
cases~\cite{Brady:1997fj,Hawley:2000dt,Noble:2003xx,Jin:2006gm,Kellermann:2010rt,Radice:2010rw,Wan:2011wg,Liebling:2010bn}.

All of the work reviewed above used a non-zero particle mass.  However, the massless case 
can also be considered and the current research is largely aimed at exploration of 
that sector.  Additionally, we attempt to address some issues that remained open
following Andr\'{e}asson and Rein's work, including whether there is any explanation for 
the indications of universality seen in~\cite{Inaki:2001}.
We note that for the massless model Martin-Garcia and Gundlach \cite{PhysRevD.65.084026}
considered the possibility of the existence of one-mode unstable self similar configurations that 
could serve as type II critical solutions.
Interestingly, they concluded that since there are infinitely
many matter configurations that give rise to any given static spacetime, any unstable solution 
must have an {\em infinite} number of unstable modes.  
Their argument also applies to the 
static case, which then suggests that there should be no type I behaviour in the model either.

In spherical symmetry the Vlasov equation is a PDE in 
time and three phase space dimensions.  Thus, direct numerical solution is 
costly and this fact motivated
the use of particle-based algorithms in all previous studies
excepting~\cite{stevenson}.  However,
a key deficiency of particle approaches is that the results develop a stochastic
character on a short time scale. This leads to poor convergence properties relative to 
a direct method, namely an error that is only $O(1/\sqrt{N})$, where $N$ is the number of 
particles.  With the substantial increase in computational resources over time,
direct solution techniques have become feasible and about a decade ago 
Stevenson~\cite{stevenson} implemented a finite-volume solver for 
the Vlasov PDE for the case that all particles have the same angular momentum.  
The code that we have developed is largely a continuation of his
effort and produces results that have well-behaved convergence properties
as a function of the mesh spacing.

Our numerical studies are based on two types of initial data.  The first, which we term {\em generic},
is characterized by a relatively arbitrary functional form for $f(0,x^i,p_j)$.  The second, which 
we call {\em near static}, is based on perturbations about some precisely static solution that is constructed from 
the ansatz described in~Sec.~\ref{sec:initstat}.  We perform experiments using initial conditions
of the first type for both massless and 
massive particles, but restrict attention to the massless sector for our near-static studies.  Aiming 
to unearth as much phenomenology as possible, as well as to explore the issue of universality, we 
have attempted to broadly survey the possibilities for the specific form of the initial distribution function in all
three sets of experiments.

The remainder of the paper is structured as follows.  The next section describes the equations of motion for 
the model while Sec.~\ref{sec:initstat} discusses the construction of  static solutions from the ansatz mentioned
previously.
Sec.~\ref{numerical} details 
our numerical approach, including code validation.  Sec.~\ref{results} is devoted to the main results from our study and we 
conclude with a summary and discussion in Sec.~\ref{conclusions}.  We have adopted units in which $G=c=1$.

\section{Equations of Motion\label{eom}}
A configuration of a system of particles can be described by the phase space density,
$f(t,x^i,p_j)$, also known as the distribution function, where $x^i$ and $p_j$ are the particles' spatial positions and 3-momenta, respectively. 
In the Einstein-Vlasov system particles interact only through  
gravity. Consequently, the particles move on geodesics of the spacetime along which 
the density function is conserved:
\begin{equation}
\frac{Df(t,x^j,p_j)}{d\tau} = 0 \, .
\end{equation}
Here, $\tau$ is the proper time of the particle and $D/d\tau$ is the Liouville operator:
\begin{equation}
   \frac{D}{d\tau} \equiv \DERIV{x^{\mu}}{\tau}  \frac{\partial}{\partial x^{\mu}} + \DERIV{{p_j}}{\tau}  \frac{\partial}{\partial p_{j}} \, .
\end{equation}
Using the geodesic equation
\begin{equation}
 v^{\mu}\partial_{\mu} p_{\nu} - v^{\mu} \Gamma^{\lambda}_{\mu\nu}p_{\lambda} = 0 \, ,
\end{equation}
where $v^\mu$ is the particle 4-velocity, the Vlasov equation can be written as
\be
p^{\mu}\frac{\partial f}{\partial x^{\mu}} + p^{\nu}p_{\lambda}\Gamma^{\lambda}_{\nu j}\frac{\partial f}{\partial p_{j}} = 0 \, .
\label{eq:vlasov}
\ee

The energy momentum tensor of the system is given by integrating over the
momentum of the particles:
\begin{equation}
T_{\mu\nu}(t,x^i) = \int \frac{p_{\mu}p_{\nu}}{m}f(t,x^i,p_j)dV_{p_j} \, ,
\label{eq:emtensor}
\end{equation}
where $m$ is the particle mass.
Equations (\ref{eq:vlasov}) and (\ref{eq:emtensor}), together with Einstein's equations
\be
G_{\mu\nu} = 8\pi T_{\mu\nu} \, ,
\ee
govern the evolution of the Einstein-Vlasov system.
These equations, restricted to spherical symmetry by 
requiring 
$f(t,x^i,p^j) = f(t,R(x^i),R(p^j))$, $R \in SO(3)$ 
is the system we study numerically.

\subsection{Coordinate choice and equations for metric components\label{sec:geometry}}
We adopt polar-areal coordinates $(t,r)$ in which the spherically-symmetric metric takes the form
\be
 ds^2 = -\alpha(t,r)^2dt^2 + a(t,r)^2dr^2 + r^2d\theta^2 + r^2\sin^2\theta d\phi^2 \, .
\ee
The radial metric function $a(t,r)$ can be determined from either the Hamiltonian constraint,
\begin{equation}
\frac{a'}{a} = \frac{1-a^{2}}{2r} - \frac{ra^{2}}{2}8\pi T^{t}_{\;\;\;t} \, ,
\label{eq:ettt}
\end{equation}
where ${}'\equiv\partial/\partial r$, or from the momentum constraint,
\begin{equation}
\frac{\dot{a}}{a} = \frac{ra^{2}}{2}8\pi T^{r}_{\;\;\;t} \, ,
\label{eq:etrt}
\end{equation}
with $\dot{~}\equiv\partial/\partial t$.
The lapse function $\alpha(t,r)$ is fixed by the polar slicing-condition
\begin{equation}
\frac{\alpha'}{\alpha} = \frac{a^{2}-1}{2r} + \frac{ra^{2}}{2}8\pi T^{r}_{\;\;\;r} \, .
\label{eq:etrr}
\end{equation}
Equation~(\ref{eq:ettt}) is solved subject to the boundary condition,
\begin{equation}
   a(t,0) = 1 \, ,
   \label{bc_a}
\end{equation}
which follows from the demand of elementary flatness at the origin.  For the lapse we set
\begin{equation}
   \alpha(t,r_{\rm max}) = \frac{1}{a(t,r_{\rm max})} \, ,
   \label{bc_alpha}
\end{equation}
where $r_{\rm max}$ is the location of the outer boundary of the computational domain,
so that coordinate and proper time coincide at infinity.

The $\theta\theta$ component of Einstein's equation 
yields an additional redundant equation, and we use the degree to which it is satisfied 
as a check of our numerical results.

\subsection{The energy momentum tensor}
As noted above,
for a given distribution function, $f(t,x^i,p_j)$, the stress tensor
is computed from the momentum-space integral~(\ref{eq:emtensor}). 
With our choice of metric the volume element 
is given by
\begin{equation}
dV_{p_j} = \frac{md^{3}p_j}{p^{0}\sqrt{|g|}} = \frac{mdp_{r}dp_{\theta}dp_{\phi}}{p^{0}\alpha a r^{2} \sin \theta} \, .
\end{equation}
To impose spherical symmetry we require the distribution function to be 
uniform in all possible angular directions.
This condition can be conveniently implemented by transforming to variables $l^2$ and 
$\psi$ given by
\begin{equation}
l^{2} \equiv p_{\theta}^{2} + \frac{p_{\phi}^2}{\sin^{2}\theta} \, ,
\end{equation}
\begin{equation}
\psi \equiv \tan^{-1}\left(\frac{p_{\theta}\sin\theta}{p_{\phi}}\right) \, ,
\end{equation}
where $l$ is the angular momentum of the particles. 
Spherical symmetry 
is then achieved by demanding that 
$f(t,x^i,p_r,l^2,\psi) \equiv f(t,r,\theta,\phi,p_r,l^2,\psi) = f(t,r,p_r,l^2)$.
The volume element in the new variables is
\begin{equation}
dV_{p_j} = \frac{mdp_{r}dl^{2}d\psi}{2a\bar{p}^{t} r^{2}} \, ,
\end{equation}
where
\begin{equation}
\bar{p}^{t}\equiv \alpha p^{0} = \sqrt{m^{2} + \frac{p_{r}^{2}}{a^{2}} + \frac{l^{2}}{r^{2}}} \, .
\label{eq:defpbart}
\end{equation}

Integrating over $\psi$, the components of the energy momentum tensor are given by:
\be
T^{t}_{\;\;\;t} = \frac{-\pi}{ar^{2}} \iint \bar{p}^{t}fdp_{r}dl^2 \, ,\\
\label{eq:ttt}
\ee
\be
T^{r}_{\;\;\;r} = \frac{\pi}{a^{3}r^{2}} \iint \frac{p_{r}^{2}}{\bar{p}^{t}}fdp_{r}dl^2 \, ,\\
\label{eq:trr}
\ee
\be
T^{r}_{\;\;\;t} = \frac{-\pi\alpha}{a^{3}r^{2}} \iint p_{r}fdp_{r}dl^2 \, ,
\label{eq:trt}
\ee
\begin{equation}
T^\theta_{\;\;\theta} = \frac{-\pi}{2ar^{4}}\iint  \frac{l^2f}{\bar{p}^{t}}dp_{r}dl^2 \, .
\label{eq:tthth}
\end{equation}

\subsection{Evolution of the distribution function}
Having imposed spherical symmetry
the Vlasov equation~(\ref{eq:vlasov}) can be written as
\begin{equation}
p^{t}\PDERIV{f}{t} + p^{r}\PDERIV{f}{r} +  
\left(\frac{\alpha'p_{t}^2}{\alpha^{3}} + \frac{a'p_{r}^{2}}{a^{3}p^{t}} + \frac{l^{2}}{r^{3}}\right) \PDERIV{f}{p_{r}} = 0 \, .
\label{eq:vlasovss}
\end{equation}
By defining
\begin{eqnarray}
g&\equiv&\frac{\alpha p_{r}}{\alpha^{2}\bar{p}^{t}}= \frac{\partial H}{\partial p_r} \, ,\\ 
h&\equiv&-\alpha'\bar{p}^{t} + \frac{\alpha a' p_{r}^{2}}{a^{3}\bar{p}^{t}} + \frac{\alpha l^{2}}{r^{3}\bar{p}^{t}} = -\frac{\partial H}{\partial r} \, ,
\end{eqnarray}
where $H$ is the Hamiltonian,
\be
H \equiv \alpha\sqrt{m^2 + (p_r/a)^2 + (l/r)^2} \, ,
\ee
equation~(\ref{eq:vlasovss}) can be cast as a conservation law:
\begin{equation}
\PDERIV{f}{t} - \{H,f\} = \PDERIV{f}{t} + \PDERIV{(gf)}{r} +  \PDERIV{(hf)}{p_{r}} = 0 \, .
\label{eq:vlasov_conserved}
\end{equation}
This form of the Vlasov equation facilitates the use of finite-volume techniques in our 
numerical treatment of the problem.

\section{Static Solutions\label{sec:initstat}}
Spherically symmetric static solutions of the Vlasov equation can be generated 
by simply requiring that the
distribution function at the initial time take the form $f(0,r,p_r,l^2) = \Phi(E,l)$,
where 
\begin{equation}
E\equiv\alpha\sqrt{m^2+(p_r/a)^2+(l/r)^2}
\end{equation}
is the energy of the particles 
and, again, $l$ is the angular momentum parameter~\cite{Rein:1993ixp}.
Indeed, since $E$ and $l$ are
both conserved along particle geodesics in spherical symmetry, any distribution function of this 
form remains unchanged as the particles move
and the Vlasov equation is automatically satisfied. 

Explicit construction of the static spacetime resulting from a given choice of $\Phi(E,l)$ 
requires that the metric functions $\alpha$ and $a$ be determined self-consistently. 
To that end we can write~(\ref{eq:ettt}) and (\ref{eq:etrr}) as
\begin{equation}
\frac{-2r\partial_r\ln a + 1}{a^2} -1 = 8\pi r^2 T^t_{\;\;t}(r;\alpha,\Phi) \, ,
\label{eq:a_func}
\end{equation}
\be
\frac{2r\partial_r\ln \alpha + 1}{a^2} -1 = 8\pi r^2 T^r_{\;\;r}(r;\alpha,\Phi) \, ,
\label{eq:alpha_func}
\ee
where
\be
T^t_{\;\;t}(r;\alpha,\Phi) = -\frac{\pi}{r^2} \iint \bar{p}^t \Phi(E(\alpha,r,w,l),l)dw \, dl^2 \, ,
\label{eq:Ttt_func}
\ee
\be
T^r_{\;\;r}(r;\alpha,\Phi) =  \frac{\pi}{r^2} \iint \frac{w^2}{\bar{p}^t} \Phi(E(\alpha,r,w,l),l)dw \,  dl^2 \, ,
\label{eq:Trr_func}
\ee
\be
w = \frac{p_r}{a} \, ,
\ee
\be 
\bar{p}^t = \sqrt{m^2+w^2+(l/r)^2} \,  ,
\ee
\begin{equation}
E=\alpha\sqrt{m^2+w^2+(l/r)^2}  \, .
\label{eq:energy}
\end{equation}
Given a functional form for $\Phi(E,l)$, we can integrate the equations for $\alpha(r)$ and 
$a(r)$ from $r=0$ outward, subject to the boundary conditions~(\ref{bc_a})-(\ref{bc_alpha}).
Physically, we also want the particle distribution resulting from 
a given $\Phi(E,l)$ to have 
compact support in phase space and finite total mass.
As shown in \cite{Rein1992}, these conditions can be satisfied by
introducing a maximum (cut-off) energy, $E_0$, so that\\
\be
\Phi(E,l) = \phi(E/E_0)\Theta(E_0-E) F(l) \, ,
\ee
where $\Theta$ is the unit step function.  
In Sec.~\ref{sec:staticmassless} we construct static solutions based on this ansatz and then investigate 
their relationship to critical behaviour in the model.

\section{Numerical Techniques\label{numerical}}
In this section we summarize our numerical approach for constructing approximate solutions of 
the equations of motion and the various tests we have performed to establish the 
correctness and accuracy of our implementation. 

\subsection{Evolution scheme}
As previously mentioned, we treat the matter evolution by a direct discretization
of the multidimensional Vlasov equation.  Relative to the particle methods 
adopted in most previous studies of the Einstein-Vlasov system, this has the advantage that 
our numerical solutions have superior convergence properties. In particular,
in contrast to the particle approach,
there is no stochastic component of the solution error.
This in turn leads to improved
confidence in our identification of key aspects of the critical phenomena
exhibited in the model, including 1) evidence that the threshold solutions
{\em are} static and 2) the scaling exponents associated with the critical
configurations.
%
%

As also noted above, the Vlasov equation can be expressed in 
conservation form and 
is thus amenable to solution using finite-volume methods.  
These techniques, which are used extensively in fluid dynamics, for example, 
are well known for their ability to accurately resolve sharp 
features---including discontinuities---that often appear in the solution 
of conservation laws.  In our case, evolutions of the distribution
function generically exhibit significant mixing and steep gradients; moreover,
some of our computations involve initial data which is not smooth in 
phase space.  The finite-volume strategy is thus natural for our purposes.
\begin{figure}
\begin{center}
\includegraphics[width=0.50\textwidth]{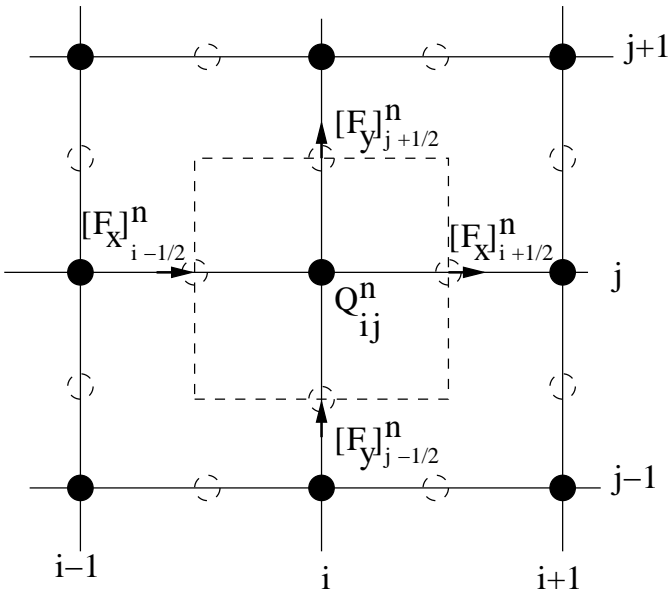}
\caption[Finite Volume Cell]{\label{fig:finite_volume_cell}
A portion of the discretized computation domain used in 
our finite volume code.  The dashed lines delineate one finite volume cell. The 
cell-centred average value of the density, $Q^n_{ij}$ is defined on the grid points 
marked with filled circles while the fluxes, $[F_x]^n_{i-1/2}$, $[F_x]^n_{i+1/2}$, etc. 
are computed at points denoted with dashed circles and which lie on cell boundaries. 
As described in more detail in the text, $Q^n_{ij}$ is updated using the difference of the outgoing and 
ingoing fluxes through the cell boundaries.
}
\end{center}
\end{figure}
We sketch our specific
approach by considering the general form of a conservation equation for 
a quantity $q(t,x,y)$:
\be
\PDERIV{q(t,x,y)}{t} + \PDERIV{F_x(q)}{x} + \PDERIV{F_y(q)}{y} = 0 \, ,
\label{eq:flux-cons}
\ee
where $F_x(q)$ and $F_y(q)$ are the fluxes
in the $x$ and $y$ directions.  
We follow the usual finite volume approach 
(see~\cite{leveque_book} for example) by dividing 
the computational domain into $N_x \times N_y$ 
cells of uniform size $\Delta x \times \Delta y$ as shown in Fig.~\ref{fig:finite_volume_cell},
and define the average value of the
unknown $q$ over the cell $C_{ij}$ by
\be
Q^{n}_{ij} = \frac{1}{\Delta x \Delta y} \iint_{C_{ij}} q(t^n,x,y) dx dy \, .
\ee
Here the superscript $n$ labels the discrete time, $t^n \equiv n\Delta t$.
We then rewrite (\ref{eq:flux-cons}) in integral form:
\bean
\frac{\partial Q}{\partial t} = - \frac{1}{\Delta x \Delta y}\left(\int_{\rm E} F_x(q) dy -  \int_{\rm W} F_x(q) dy\right) \nonumber\\
- \frac{1}{\Delta x \Delta y}\left(\int_{\rm N} F_y(q) dx - \int_{\rm S} F_y(q) dx\right) \, ,
\eean
where the subscripts E, W, N and S denote the east, west, north and south
boundaries, respectively, of the cell $C_{ij}$.
Applying a time-discretization to this last expression yields an equation 
that can be used to advance the cell average in time:
\bean
Q_{ij}^{n+1} = Q_{ij}^n - \frac{\Delta t}{\Delta x}\left(
\left[F_x\right]^n_{i+1/2}- \left[F_x\right]^n_{i-1/2}\right)  \nonumber \\
-
\frac{\Delta t}{\Delta y}\left(
\left[F_y\right]^n_{j+1/2}- \left[F_y\right]^n_{j-1/2}
\right) \, .
\label{eq:timestep_vlasov}
\eean
Here the average fluxes at the boundaries, $\left[F_x\right]^n_{i+1/2}$ etc. are 
calculated using a Roe solver~\cite{leveque_book}.   We note that our calculations 
are always performed on meshes that are uniform in each coordinate direction, and 
that when we change resolution---to perform a convergence test for example---each 
mesh spacing is changed by the same factor.  Thus, our discretization is fundamentally
characterized by a single scale, $h$.  Our specific finite volume approach is 
based on $O(h^2)$ approximations.  However, the nature of the flux 
calculations---which are designed to inhibit the development of spurious oscillations---means 
that the scheme is only $O(h)$ in the vicinity of any local extrema in the solution.


The metric variables $\alpha$ and $a$, which need only be defined on a mesh in the $r$ 
direction, are computed
from $O(h^2)$ finite difference approximations of the Hamiltonian and slicing
equations, (\ref{eq:ettt}) and (\ref{eq:etrr}).  Since the equations for the matter and 
geometry are fully coupled---i.e. $\alpha$ and $a$ appear in the flux computations, and 
$f$ is needed for the calculation of the source terms for $\alpha$ and $a$---some 
care is needed to construct a scheme which is fully $O(h^2)$ accurate (modulo the 
degradation of convergence near extremal solution values just noted).
In practice, we use an $O(\Delta t^2)=O(h^2)$ 
Runge-Kutta scheme for the time stepping, which necessitates computation of auxiliary 
quantities at the half time step $t^{n+1/2} = t^n + \Delta t/2$.  Our overall scheme
that advances
the solution from $t^n$ to $t^{n+1}$, and which {\em does} have $O(h^2)$ truncation error, is:
\begin{enumerate}
\item Compute $f^{n+1/2}$ from (\ref{eq:timestep_vlasov}) using the fluxes $F^n$.
\item Compute ${\tilde a}^{n+1/2}$ from (\ref{eq:etrt}) with source $\left[ T^r{}_t \right]^{n}$.
\item Compute $\left[ T^t{}_t \right]^{n+1/2}$ and $\left[ T^r{}_r \right]^{n+1/2}$
   from
   (\ref{eq:ttt})--(\ref{eq:trr}) using 
   ${\tilde a}^{n+1/2}$.
\item Compute $a^{n+1/2}$ and $\alpha^{n+1/2}$ from (\ref{eq:ettt}) and (\ref{eq:etrr}) with sources 
   $\left[ T^t{}_t \right]^{n+1/2}$ and $\left[ T^r{}_r \right]^{n+1/2}$.
\item Compute $\left[ T^r{}_t \right]^{n+1/2}$ from (\ref{eq:trt}).
\item Compute fluxes $F_x^{n+1/2}$ and $F_y^{n+1/2}$ using $a^{n+1/2}$ and $\alpha^{n+1/2}$.
\item Compute $f^{n+1}$ from (\ref{eq:timestep_vlasov}) and the half-step fluxes $F^{n+1/2}$.
\item Compute ${\tilde a}^{n+1}$ from (\ref{eq:etrt}) with source $\left[ T^r{}_t \right]^{n+1/2}$.
\item Compute $\left[ T^t{}_t \right]^{n+1}$ and $\left[ T^r{}_r \right]^{n+1}$ from 
   (\ref{eq:ttt}) and (\ref{eq:trr}) using ${\tilde a}^{n+1}$.
\item Compute $a^{n+1}$ and $\alpha^{n+1}$ from (\ref{eq:ettt}) and (\ref{eq:etrr}) using sources 
   $\left[ T^t{}_t \right]^{n+1}$ and $\left[ T^r{}_r \right]^{n+1}$.
\item Compute $\left[ T^r{}_t \right]^{n+1}$ from (\ref{eq:trt}).
\item Compute fluxes $F_x^{n+1}$ and $F_y^{n+1}$ using $a^{n+1}$ and $\alpha^{n+1}$.
\item One time step complete; start next time step.
\end{enumerate}

To facilitate the use of large grid sizes, as well as to speed up the
simulations, we parallelize the computations for the evolution of the 
distribution function and the calculation of the energy-momentum tensor components using the PAMR
package~\cite{PAMR}.  On the other hand, the calculation of the metric components, which 
has negligible cost relative to the updates of $f$ and $T^\mu{}_\nu$, is performed on a single
processor.  The new values of the metric functions are then broadcast to the other CPUs.

\subsection{Initial data}
In spherical symmetry the gravitational field has no dynamics beyond that generated by 
the matter content, so 
initial conditions for our model are completely fixed by the specification of the 
initial-time particle distribution function, $f(0,r,p_r,l^2)$.  
However, the Einstein equations
(\ref{eq:ettt})--(\ref{eq:etrr}) must also be satisfied at the initial time and,
through the definition~(\ref{eq:defpbart}) for ${\bar p}^t$, $a$ appears within the integrands for the 
stress tensor components.  To determine all requisite initial values consistently we 
therefore use the following iterative scheme:
\begin{enumerate}
\item Initialize the distribution function, $f(0,r,p_r,l^2)$, to a localized function on phase space.
\item Initialize the geometry to flat spacetime.
\item Calculate the energy momentum tensor using the current geometry.
\item Calculate the geometry using the current energy momentum tensor.
\item Iterate over the matter and geometry calculations until a certain tolerance is achieved.
\end{enumerate}
In practice we find that this algorithm converges in a few iterations.

As discussed in Sec.~\ref{sec:staticmassless}, when we study static initial data we first specify 
$\Phi(E,l)$ and then integrate~(\ref{eq:a_func})--(\ref{eq:alpha_func}) outward.  
We note that the form of $\Phi(E,l)$ that we choose, 
\be
\Phi(E,l)=\phi(E/E_0)\Theta(E_0-E)F(l) \, ,
\label{eq:staticconf}
\ee
results in equations that are invariant under the transformation:
\begin{eqnarray}
   \alpha &\rightarrow& k\alpha\, , \\
   E_0 &\rightarrow& kE_0 \, .
\label{eq:escaling}
\end{eqnarray}
We can thus first integrate the slicing condition~(\ref{eq:alpha_func}) subject
to the boundary condition, $\alpha(0,0)=\Lambda$, with $\Lambda<1$ but otherwise arbitrary, and then linearly rescale $\alpha(0,r)$ 
so that $\alpha(0,r_{\rm max}) = 1 / a(0,r_{\rm max})$.  The 
central redshift of the configuration,  $Z_c$, which we use in our analysis below, is then given by
\begin{equation}
Z_c \equiv \frac{1}{\alpha(0,0)} - 1 \, ,
\label{eq:defredshift}
\end{equation}
where $\alpha(0,0)$ is now the rescaled value.
It is important to emphasize that different choices for $\Lambda$ result in distinct solutions, so 
that irrespective of any adjustable parameters that may appear in the specification of $\phi$, 
equation~(\ref{eq:staticconf}) will always implicitly define an entire family of static configurations.

\subsection{ Diagnostic quantities and numerical tests }

We have validated our implementations of the algorithms described above using a standard 
convergence testing methodology that examines the behaviour of the numerical solutions as a function
of the mesh spacing, $h$, keeping the initial data fixed.  This section summarizes the tests 
we perform---which involve
derived quantities that should be conserved in the continuum limit as well as 
the full solutions themselves---and 
presents results from their application to a representative initial data set using three
scales of discretization, $h$, $h/2$ and $h/4$.
\subsubsection{Conserved quantities}
\begin{figure}
\begin{center}
\includegraphics[width=0.50\textwidth]{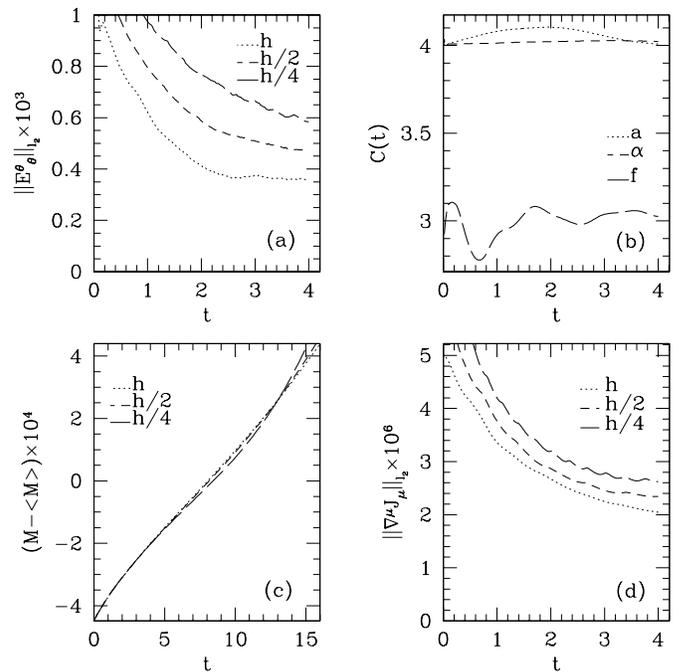}
\caption[Numerical Tests]{\label{fig:tests}  
Results of various diagnostic tests used to test the numerical solver.
%
%
The initial data and mesh resolutions used here are typical of any of the 
2D calculations described in the paper.  
A standard convergence testing methodology,
using three calculations with fixed initial data and mesh spacings 
$h$, $h/2$ and $h/4$, 
is employed. The coarsest mesh has 
$n_x \times n_y = n_r \times n_p = 128 \times 128$ grid points.
Plots (a), (c) and (d) all display quantities that are residual in 
nature, i.e. which should tend to zero quadratically in the mesh spacing.  
Values from the $h/2$ and $h/4$ computations
have been rescaled by factors of $4$ and $16$, respectively, and 
the near-coincidence of the rescaled values 
thus demonstrates that all three quantities are converging at the expected $O(h^2)$ rate.
(a)~Convergence of the $l_2$ norm of the
independent residual, $\Vert E^\theta{}_\theta \Vert_2$, defined by (\ref{eq:resthth}).
(b)~Convergence factors~(\ref{eq:convf}) of 
the primary dynamical unknowns.
Here, convergence of the metric functions, $\alpha$ and $a$, is clearly second order, while 
that for the distribution function is better than $O(h)$ but is not $O(h^2)$.  This latter 
behaviour is to be expected since the finite volume method used to update $f$ is only 
first order in the vicinity of local extrema.
(c)~Convergence of the deviation in computed total mass, calculated from~(\ref{mass_aspect}) 
and~(\ref{eq:mass}).
(d)~Convergence of the particle 
flux divergence~(\ref{eq:resid_J}).
}
\end{center}
\end{figure}

The mass aspect function, $m(t,r)$, is given by 
\be
m(t,r) = \frac{r}{2}\left(1-\frac{1}{a^{2}(t,r)}\right) \, ,
\label{mass_aspect}
\ee
and measures the amount of mass contained within radius $r$ at time $t$. Its value at
spatial infinity
\be
M \equiv m(t,\infty) \, ,
\label{eq:mass}
\ee
is the conserved ADM mass.
Alternatively, $M$ can be computed using 
\begin{eqnarray}
   M &=& \int_{0}^{\infty} \rho 4\pi r^{2} dr \, , \\
   \rho&=&n^{\mu}n^{\nu}T_{\mu\nu} \, ,
\end{eqnarray}
where $n^{\mu}$ is the unit timelike vector normal to the spatial slices. 
In developing our code we computed mass estimates based on both of these expressions, 
but the results presented here and in the remainder of the paper use~(\ref{eq:mass}) exclusively.
Fig.~\ref{fig:tests}(c) graphs deviations of $M$ relative to its time-averaged mean 
value $\langle M \rangle$ for the three computations performed with mesh scales
$h$, $h/2$ and $h/4$.  As noted in the caption, the values of $M-\langle M \rangle$
have been rescaled such that the near coincidence of the plots signals the 
expected $O(h^2)$ convergence to conservation.

The second conserved quantity that we monitor is the real-space particle flux,
$J_{\mu}$, given by
\begin{equation}
J_{\mu}(t,r) = g_{\mu\nu}\iint \frac{p^{\nu}}{m} fdV_{p_j} \, .
\end{equation}
In spherical symmetry, the only nonzero components of $J_\mu$ are
\begin{equation}
J_{t} = -\frac{\alpha \pi}{ar^{2}} \iint f(t,r,p_{r})dp_{r} \, , \\ 
\end{equation}
\begin{equation}
J_{r} = \frac{\pi}{ar^{2}} \iint \frac{p_{r}}{\bar{p}^{t}} f(t,r,p_{r})dp_{r} \, .\\ 
\end{equation}
The divergence of the flux must remain zero as the system evolves---written explicitly we have
\begin{eqnarray}
   \grad^{\mu}J_{\mu} &=& \frac{1}{\alpha^{3} a^{3} r} \Big(  - a^{3}r\dot{J_{t}}\alpha + 
         a^{3}rJ_{t}\dot{\alpha} + arJ_{r} \alpha^{2} \alpha' \nonumber \\
         + \, \,\alpha^{3}rJ_{r}'a 
         \!&-&\! \alpha r J_{t} a^{2} \dot{a} - \alpha^{3} r J_{r} \alpha' 
       + 2J_{r} \alpha^{3} a \;\; \Big) = 0 \, .
\label{eq:resid_J}
\end{eqnarray}
Plots of the rescaled $\ell_2$ spatial norm of $\grad^{\mu}J_{\mu}$ as a function of time are shown in Fig.~\ref{fig:tests}(d)---again
$O(h^2)$ convergence is observed.
\subsubsection{Independent residual test}
As noted in Sec.~{\ref{sec:geometry}}, the $\theta\theta$ component of Einstein's equation is not used in 
our evolution scheme but must be satisfied in the continuum limit if our numerical results
are valid.  We thus define the residual
\begin{equation}
	E^\theta{}_\theta \equiv G^\theta{}_\theta - 8 \pi T^\theta{}_\theta \, ,
  \label{eq:resthth}
\end{equation}
where 
\begin{eqnarray}
G^{\theta}_{\;\;\theta} &=&  G^{\phi}_{\;\;\phi} \nonumber \\
&=&-\frac{1}{r \alpha^{3} a^{3}} \left(  -\alpha^{2} a \alphaprime + \alpha^{3} \aprime + \alpha^{2} r \alphaprime \aprime \right. \nonumber \\
&-& \left. \alpha^{2} a r \frac{\partial^{2} \alpha}{\partial r^{2}}   +   a^{2}\alpha r \frac{\partial^{2} a}{\partial t^{2}}   - a^{2} r \alphadot \adot   \right) \, ,
\end{eqnarray}
and $T^\theta{}_\theta$ is given by~(\ref{eq:tthth}).
Then, using second-order finite differences to approximate all derivatives, we monitor the $\ell_2$ norm 
of $E^\theta{}_\theta$ during the calculations.   
We expect $\Vert E^\theta{}_\theta \Vert_2$ to be $O(h^2)$ and Fig~\ref{fig:tests}(a) shows that this is the case.

\subsubsection{Full-solution convergence test}
The final check we perform is a basic convergence test of
the primary dynamical variables, $\alpha$, $a$ and $f$.  Denoting the 
values computed at resolution $h$ for any of these by $q^h(t,X)$---where 
$X=r$ for $\alpha$ and $a$, and $X=(r,p_r)$ for $f$---we calculate convergence
factors, $C(t;q)$, defined by 
\be
C(t;q) = \frac{||q^h(t,X)-q^{h/2}(t,X)||_{l_2}}{||q^{h/2}(t,X)-q^{h/4}(t,X)||_{l_2}} \, .
\label{eq:convf}
\ee
If our scheme is $O(h^2)$ convergent then it is easy to argue that $C(t;q)$ should approach
4 in the continuum limit.  Plots of $C(t;a)$, $C(t,\alpha)$ and $C(t;f)$ are shown in 
Fig.~\ref{fig:tests}(b).  Second order convergence of the geometric variables is apparent,
while the behaviour of $C(t;f)$ reflects the fact that the finite
volume method we use is only first-order accurate in the vicinity of extrema of $f$.
Interestingly, at least at the resolutions used here, the deterioration of the convergence 
of $f$ does not appear to significantly impact that of the geometric quantities.

\section{Results\label{results}}

\begin{table}
\begin{center}
{\footnotesize
\begin{tabular}{| c | c | c | c |}
 \hline
  Family & $D$ & $f(0,r,p_r,l)$ & $p$  \\ \hline
  G1   & 2 & $\delta(l-l_0)\mathcal{G}(A,r_c,p_c)$ & $p_c$  \\ \hline
  G2 & 2 & $\delta(l-l_0)\mathcal{G}(A,r_c,p_c)$ & $l_0$  \\ \hline
  G3  & 2 & $\delta(l-l_0)\mathcal{G}(A,r_c,0)$ & $A$  \\ \hline
  G4 & 2 & $\delta(l-l_0)\left(\mathcal{G}(A,r_c,p_c) + \mathcal{G}(A,r_c+\Delta r,p_c+\Delta p)\right)$ & $p_c$ \\ \hline
  G5 & 2 & $\delta(l-l_0)\mathcal{E}(A,r_c,p_c)$ & $p_c$  \\ \hline
  G6 & 2 & $\delta(l-l_0)\mathcal{E}(A,r_c,0)$ & $A$  \\ \hline
  G7 & 2 & $\delta(l-l_1) \mathcal{G}(A,r_1,p_1) + \delta(l-l_2)\mathcal{G}(A,r_2,p_2)$  & $p_1$ \\ \hline
  G8 & 3 & $\exp ( -(l - l_0)^2/\Delta l ^2)\mathcal{G}(A,r_c,p_c)$ & $p_c$  \\ \hline
  G9 & 3 & $\exp ( -(l - l_0)^2/\Delta l ^2)\mathcal{G}(A,r_c,0)$ & $A$  \\ \hline
  G10 & 3 & $\Theta(l-5)\Theta(15-l)\mathcal{E}(A,r_c,0)$ & $A$  \\ \hline
\end{tabular}
}
\end{center}
\caption[Generic Families]{\label{table:generic}
Families of generic initial data used in the studies described in text.  The columns 
enumerate: (1)~the label for the family, (2)~the number, $D$,  of phase-space dimensions on which
the distribution function depends (and therefore whether the 2D or 3D code was used to generate
the results), (3)~the form of the 
initial data, $f(0,r,p_r,l)$ 
(see (\ref{eq:defG}) and (\ref{eq:defE}) for the definitions of $\mathcal{G}$ and $\mathcal{E}$), and (4)~the 
control parameter, $p$,  that was varied to study the critical behaviour.  The 
quantities $l_0, l_1, l_2, r_c, r_1, r_2, p_c, p_1, p_2, \Delta r$ and $\Delta p$ that appear 
in the various specifications of $f(0,r,p_r,l)$ are all parameters; i.e they have fixed
scalar values in any given computation.
}
\end{table}

In this section we describe the main results from our investigation of critical 
behaviour in the Einstein-Vlasov model.  We have used many different families 
of initial data in our studies and what we report below is based on a representative 
sample of those.  As mentioned in the introduction, the numerical experiments fall 
into three broad classes.  The first uses massless particles and initial data which has 
some relatively arbitrary form  
in phase space.  The second also uses massless particles but with initial 
conditions that represent perturbed static solutions.  Finally, the third 
set is the same as the first but with massive particles. We will refer to these classes
as generic massless, near-static massless, and generic massive, respectively.
In addition, the calculations can be categorized according to whether $l$ is a single 
fixed 
value, $l_0$, (2D) or if the distribution function has non-trivial $l$-dependence (3D). 
The functional form of the various families considered, along with the dimensionality 
of the corresponding PDEs and the parameter used for tuning to criticality are 
summarized in Table~\ref{table:generic}.



\subsection{Generic massless case}
\begin{figure}
\begin{center}
\includegraphics[width=0.23\textwidth]{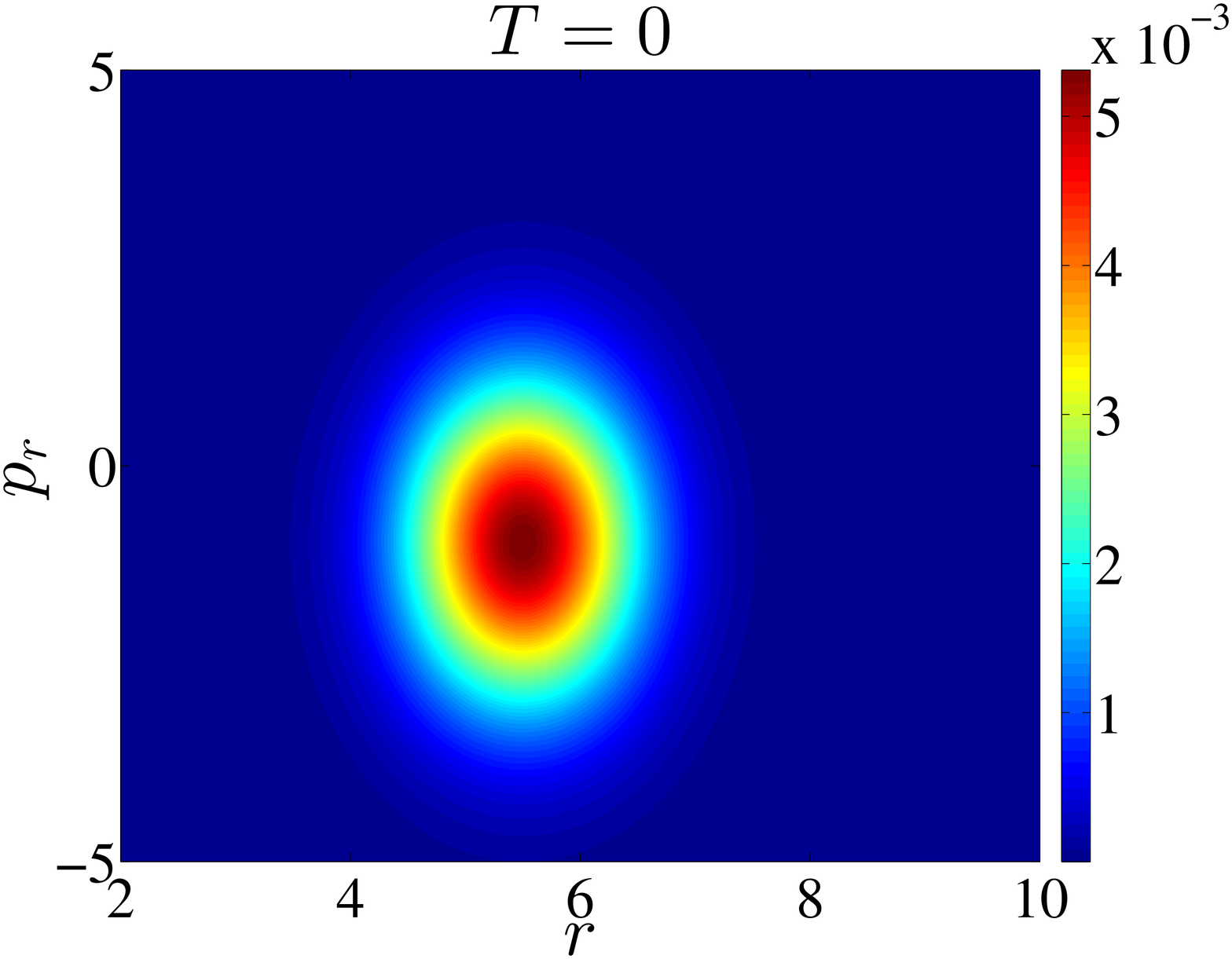}~
\includegraphics[width=0.23\textwidth]{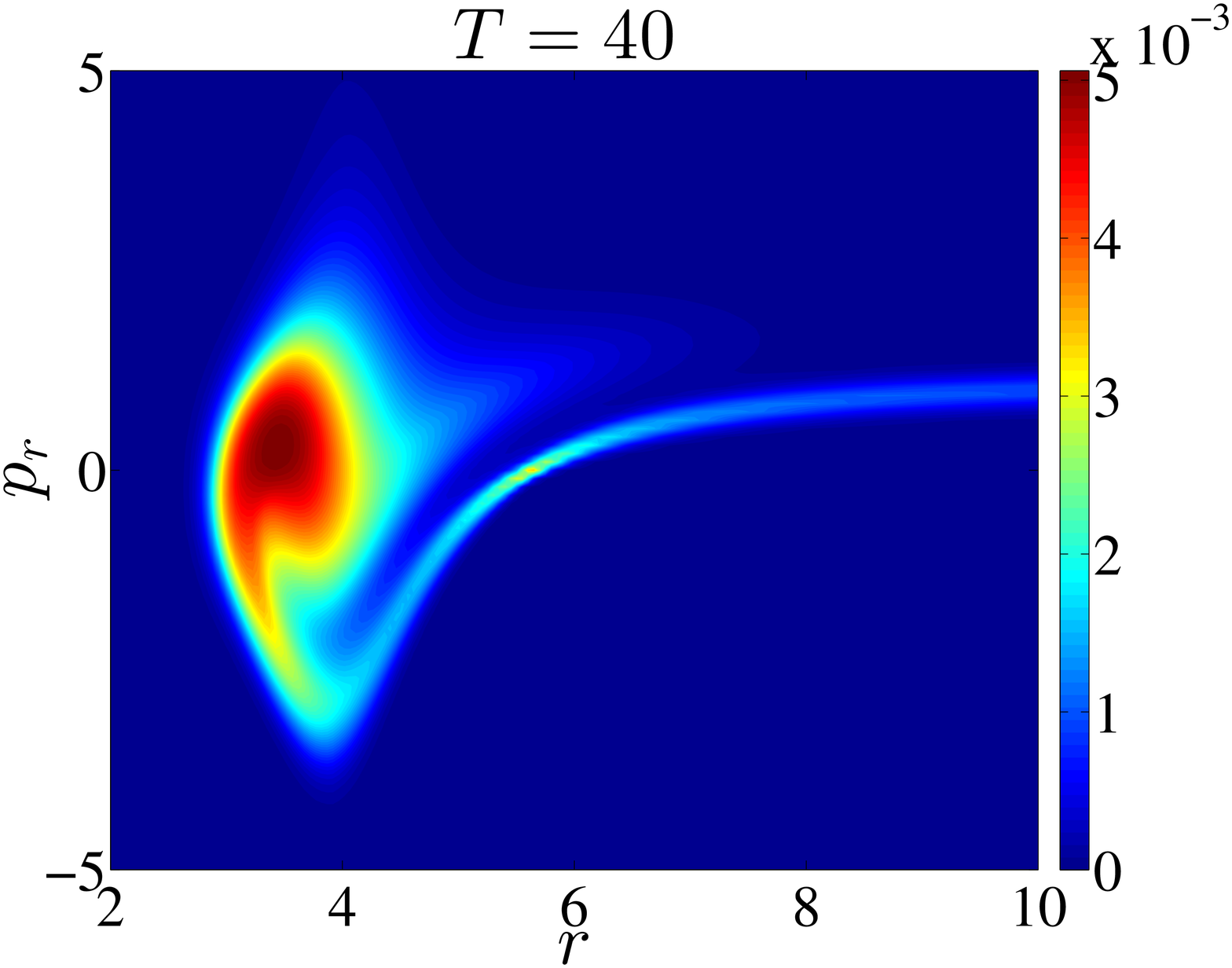}\\
\includegraphics[width=0.23\textwidth]{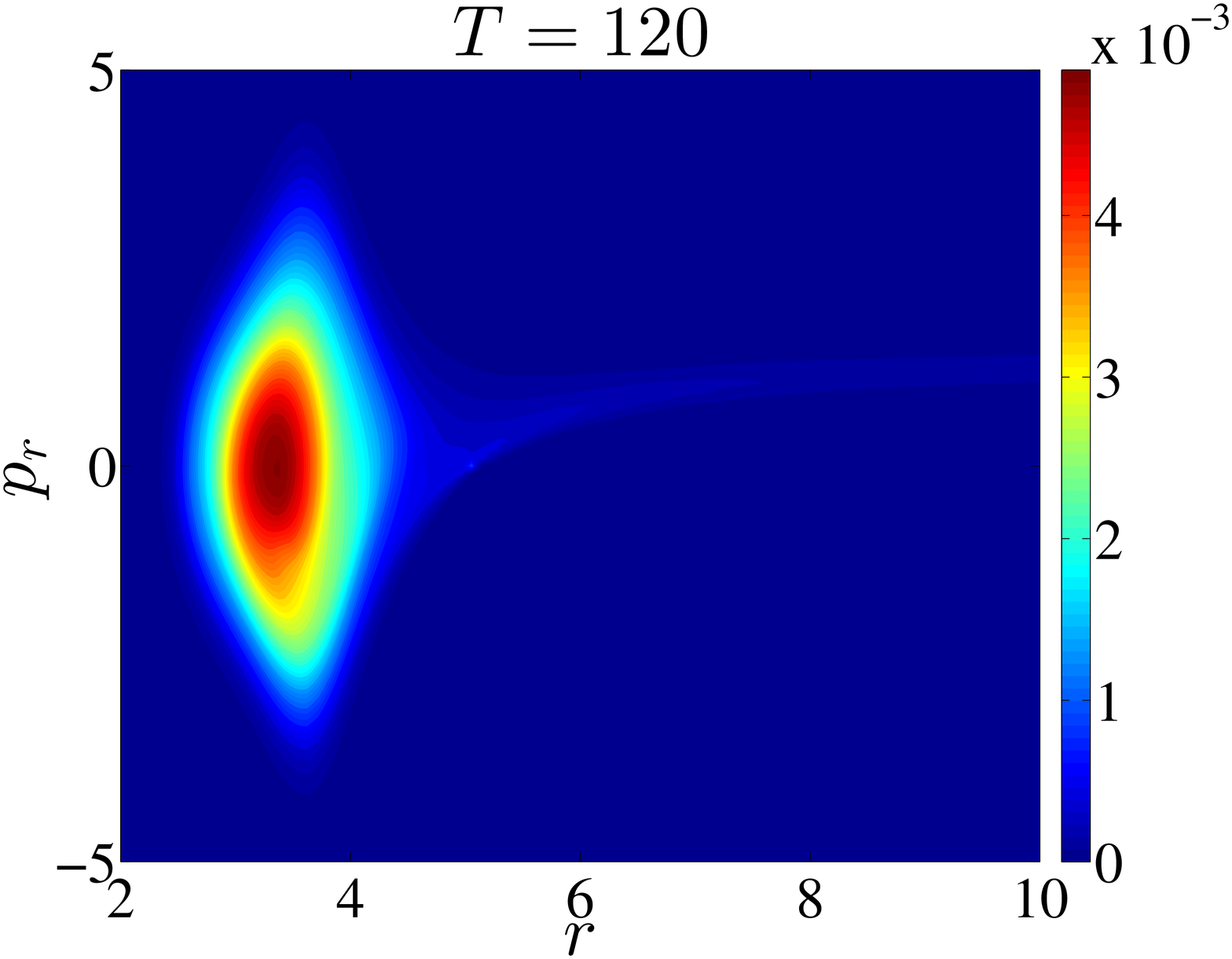}~
\includegraphics[width=0.23\textwidth]{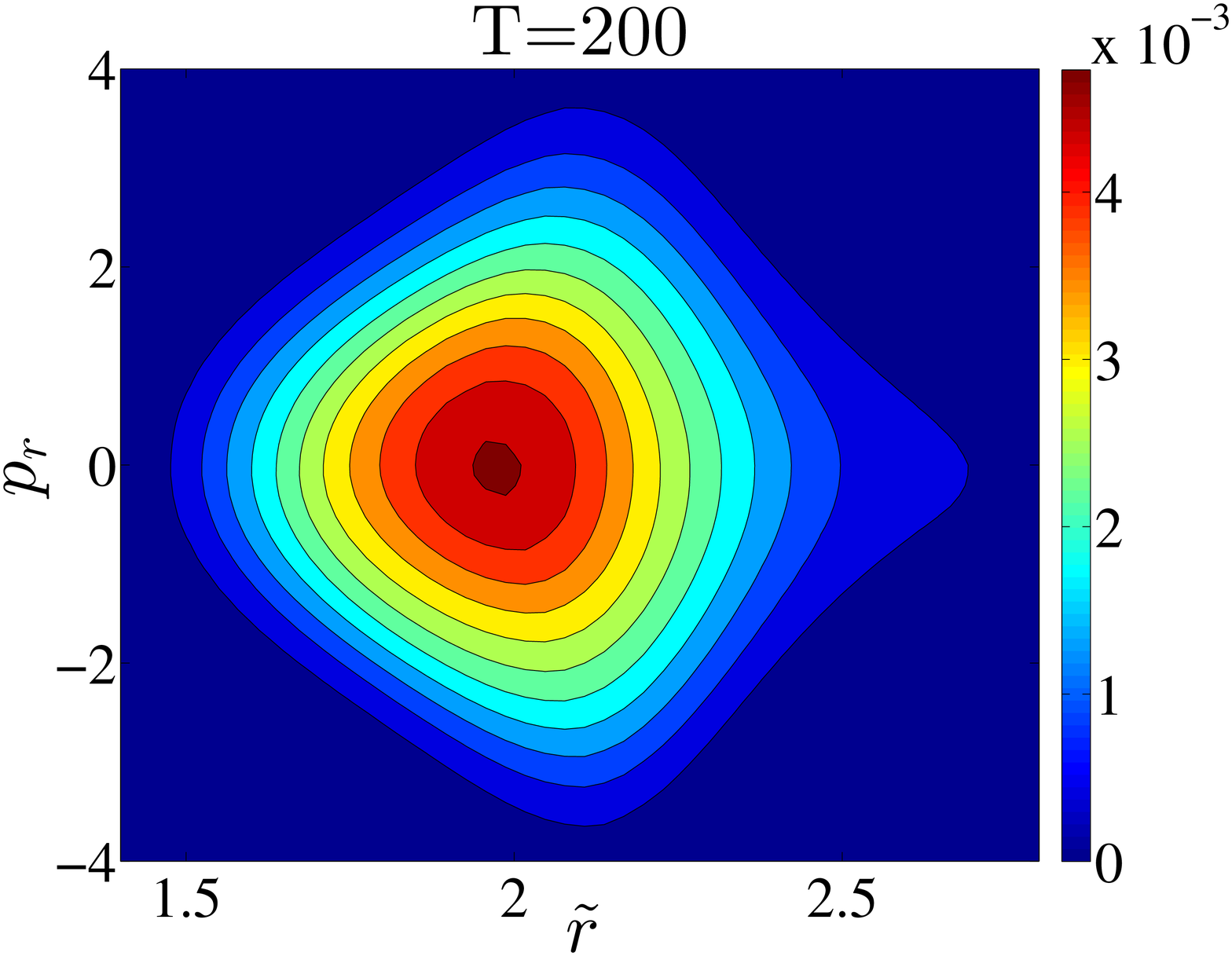}\\
\caption[Phase Space Evolution]{  \label{fig:dynamic_phasespace}  
Snapshots of the distribution function from a typical near-critical calculation, with 
evolution proceeding left to right, top to bottom (note the reduction in the 
range of radial coordinate in the last frame).
The displayed results are from family G8 (see Table~\ref{table:generic}) where $p_c$---which 
is loosely the average momentum of the initially imploding shell of particles---is the 
control parameter. As with all of the calculations discussed in the results section, the 
control parameter
has been tuned to roughly machine precision. In the early stages of the evolution we 
observe phase space mixing and the ejection of some particles (the latter particularly
visible as the ``tail'' in the second frame).  At intermediate times the system 
approaches a static state which persists for a period that is long compared to 
the infall/dispersal timescale characterizing weak field dynamics.  We note 
that this is a 3D calculation, with $f$ non-trivial in the $l$ direction: for visualization
purposes we have integrated over $l$ to produce a quantity depending only on $r$ and 
$p_r$.
Additionally, the first three frames are plotted using the computational coordinate, $r$,
while for the purposes of direct comparison with Fig.~\ref{fig:dynamic_phasespace2}, the fourth 
uses the rescaled coordinate, ${\tilde r}$, defined by~(\ref{eq:defrtilde}).
We emphasize 
that at criticality $f$ retains non-trivial dependence on $p_r$; that is, although the 
geometry is static, the particle behaviour is still dynamic.
}
\end{center}
\end{figure}

\begin{figure}
\begin{center}
\includegraphics[width=0.23\textwidth]{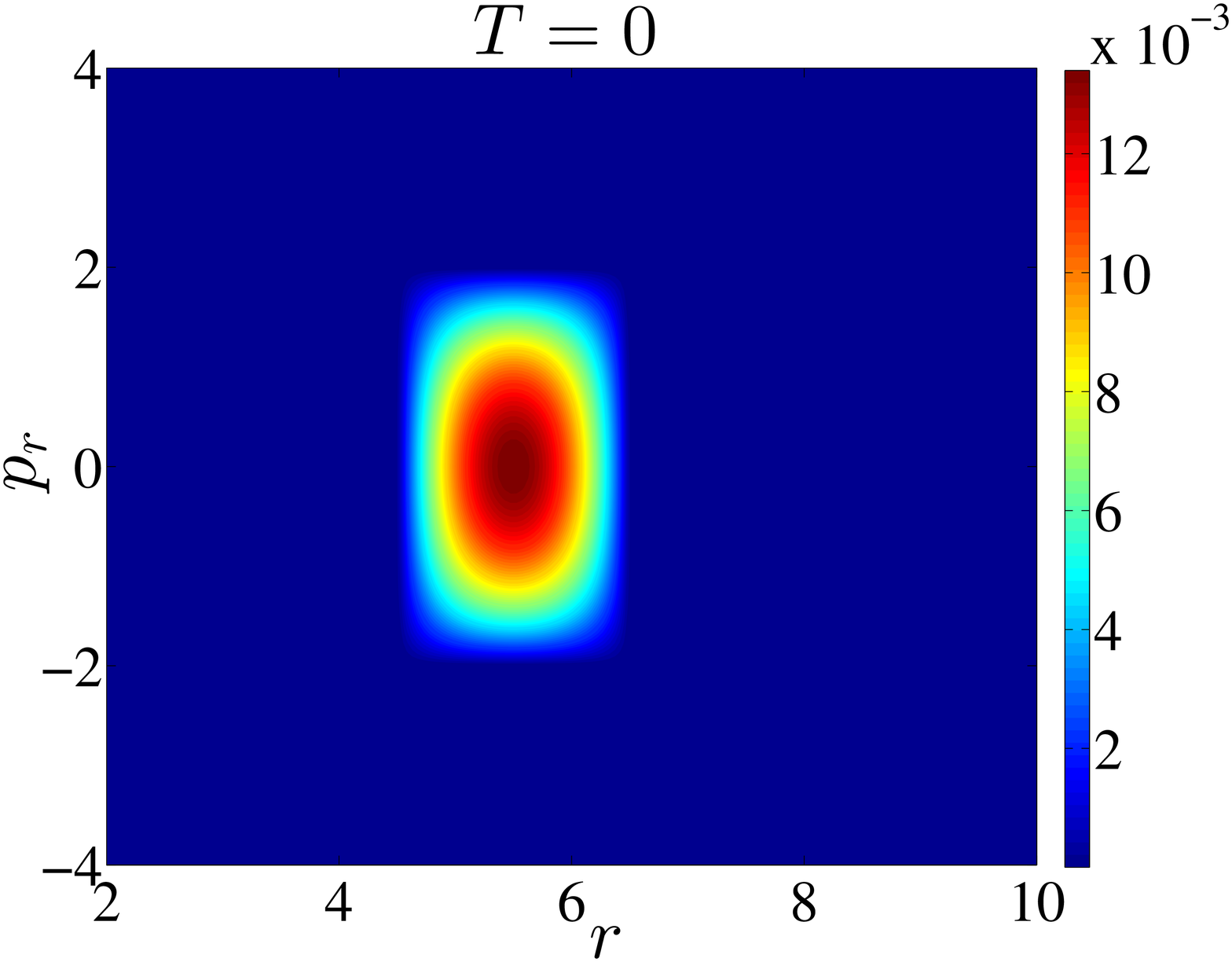}~
\includegraphics[width=0.23\textwidth]{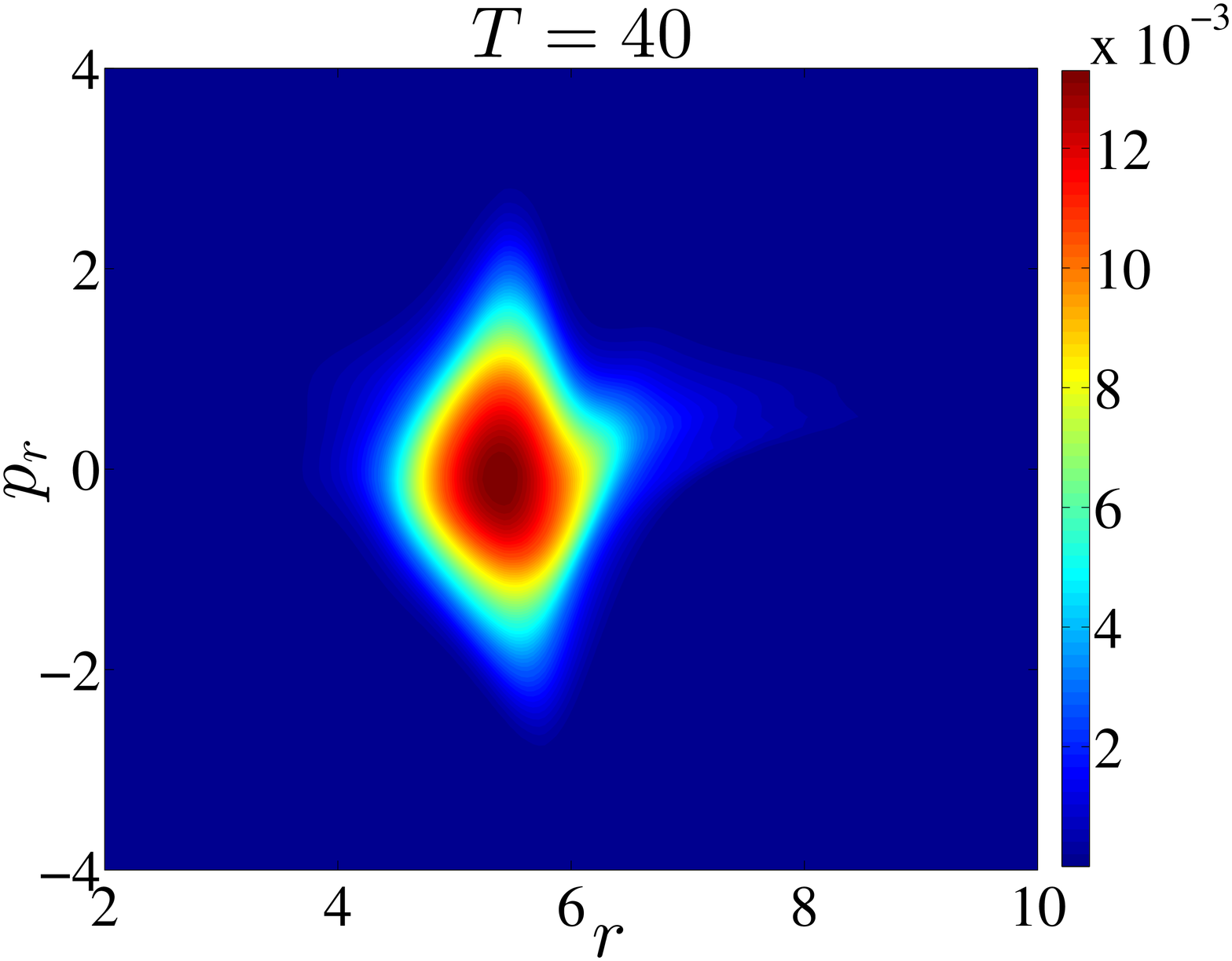}\\
\includegraphics[width=0.23\textwidth]{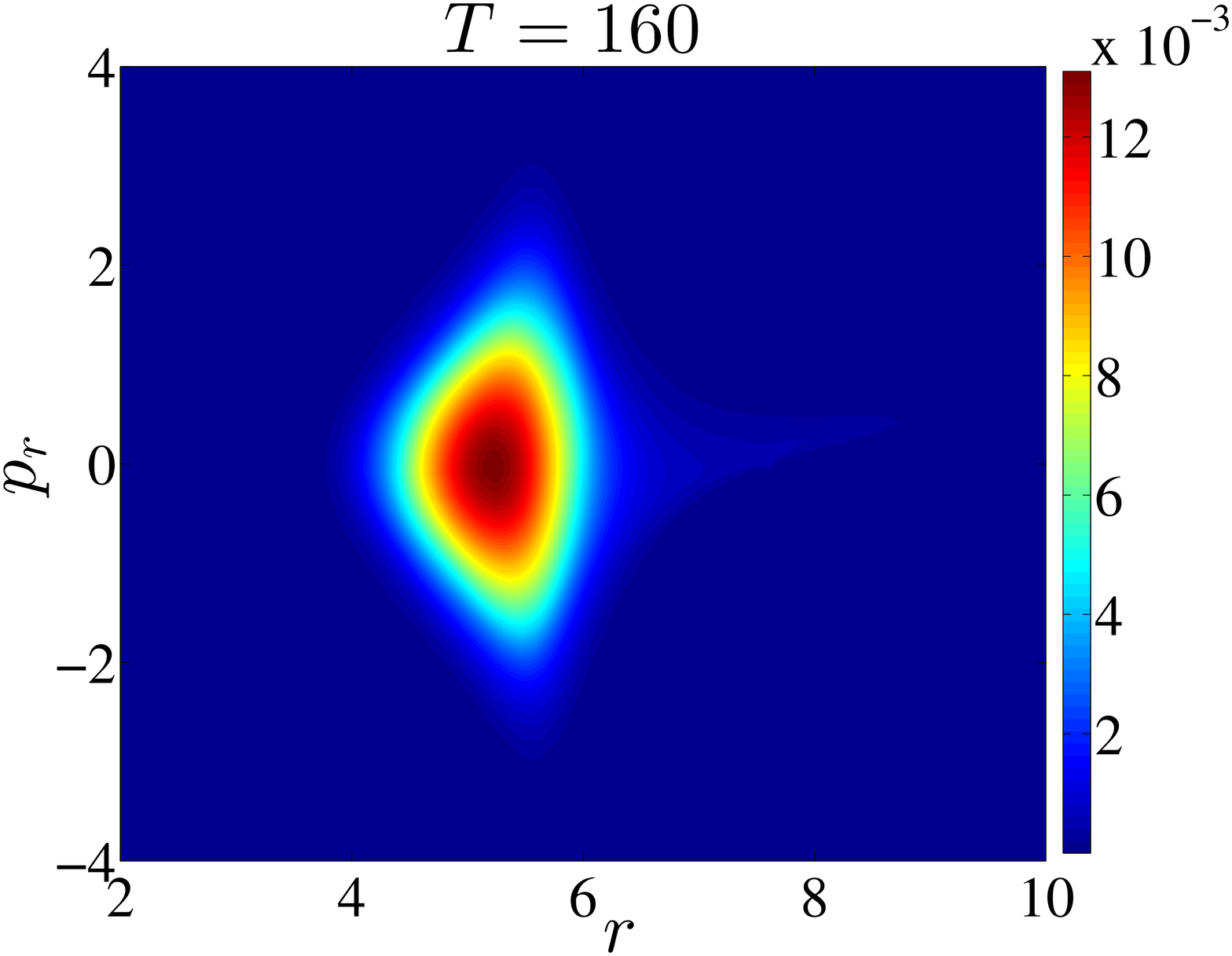}~
\includegraphics[width=0.23\textwidth]{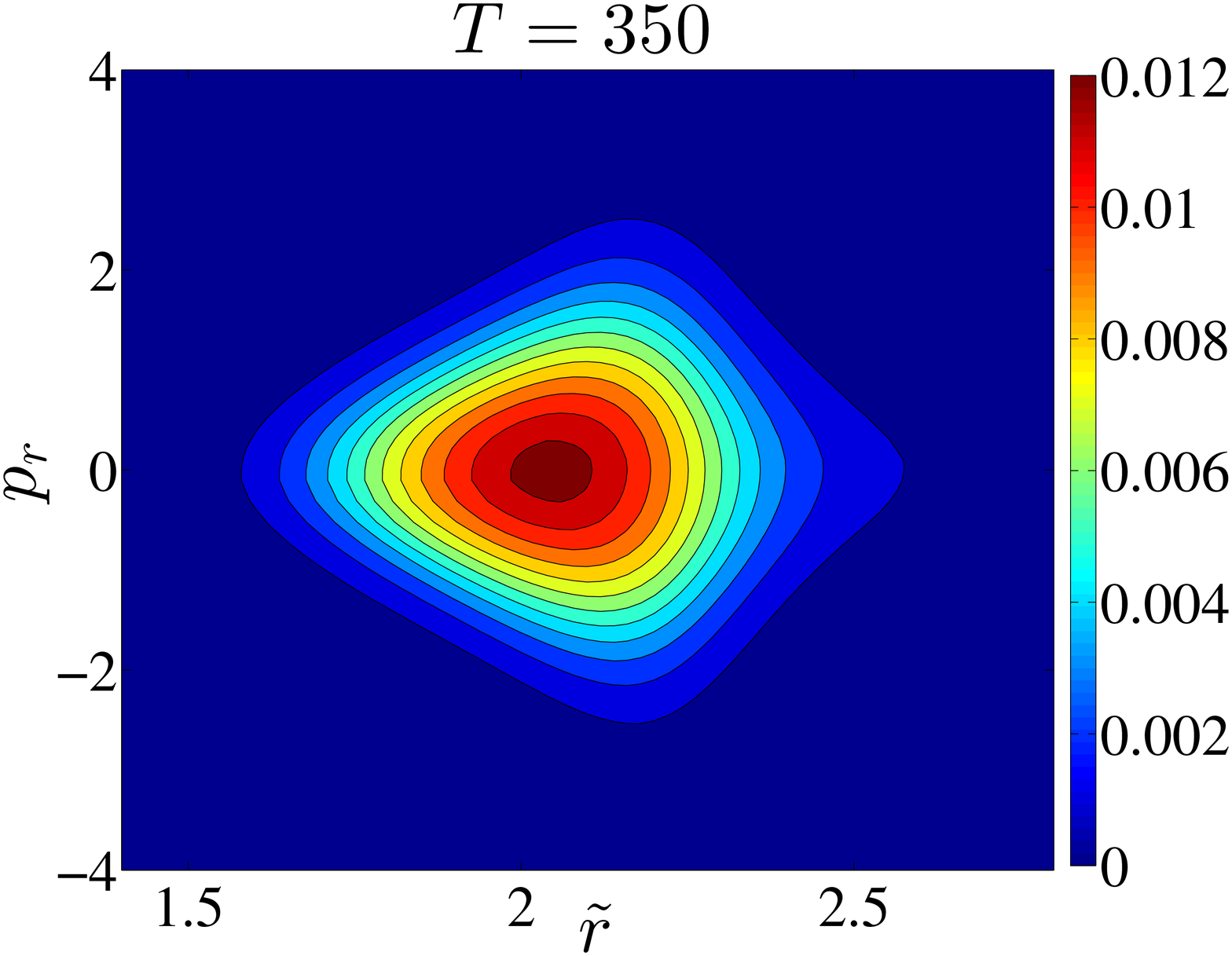}\\
\caption[Phase Space Evolution]{  \label{fig:dynamic_phasespace2}  
Snapshots of the distribution function for a near-critical calculation using 
family G10.  Here the tuning parameter is the overall amplitude, $A$, of the
initial particle distribution.  As in the previous figure the sequence shows 
an approach to a static state, but it is evident that the form of the distribution
function at criticality is significantly different in the two calculations. 
Due to the use of the rescaled 
radial coordinate, ${\tilde r}$, the fourth frames of the two figures can
be meaningfully compared.
}
\end{center}
\end{figure}


Here we use initial distribution functions, $f_0\equiv f(0,r,p_r,l)$, that describe configurations 
of particles localized in $r$, $p_r$ and $l$, and that include various parameters which 
can be tuned to generate families of solutions that span the black hole threshold.
Specifically, we set 
\begin{equation}
	f(0,r,p_r,l^2) = S(r,p_r)F(l) \, ,
\end{equation}
where $S(r,p_r)$ is given by either a gaussian function,
\be 
\mathcal{G}(r, p_r ;\, A,r_c,p_c)  \equiv A \exp \left( - \frac{(r-r_c)^2}{\Delta_r^2} - \frac{(p_r-p_c)^2}{\Delta_p^2}  \right) \, ,
\label{eq:defG}
\ee
or the truncated bi-quadratic form
\be
\mathcal{E}(r, p_r ;\, A,r_c,p_c) \equiv \left\{
  \begin{array}{l l }
  A \bar{r}(1-\bar{r})\bar{p}(1-\bar{p}) &  0<\bar{r}<1\,,\\
   ~~~~~~~~~~~~~~~~~~~~~~~~~~~ &  0<\bar{p}<1 \,,\\
    $0$ & \text{elsewhere,} 
  \end{array} \right.
\label{eq:defE}
\ee 
where $\bar{r} = (r-r_c+\Delta r)/ 2\Delta r$ and $\bar{p} = (p_r-p_c+\Delta p)/2\Delta p$.
Note that the dependence of $\mathcal{G}$ and $\mathcal{E}$ on $r$ and $p_r$ is 
suppressed in the abbreviated notation used in
Table~\ref{table:generic}.  For the 3D calculations, 
we use two types of angular momentum distribution: the first is a gaussian,
\be
F(l) = \exp\left( \frac{-(l - l_0)^2}{\Delta l ^2}\right) \, ,
\ee
while 
the second is uniform in $l$ with cutoffs at some prescribed minimum and maximum values,
$l_{\rm min}$ and $l_{\rm max}$, respectively,
\be
F(l) =  \Theta(l-l_{\rm min})\Theta(l_{\rm max}-l)  \, .
\ee


It is important to point out that since the massless 
Einstein-Vlasov system is scale-free it has an additional symmetry relative
to the massive case.  Specifically, the equations of motion are invariant
under the transformation
\begin{eqnarray}
t \rightarrow kt \, , 
\label{eq:scaletransformationt} \\
r \rightarrow kr \, ,
\label{eq:scaletransformationr}
\end{eqnarray}
where $k$ is an arbitrary positive constant.
In order to meaningfully compare results from different initial data 
choices we must therefore adopt unitless coordinates in our analysis.  We do 
this by rescaling 
$t$ and $r$ by the total mass, $M^\star$, of the putatively static solution
which arises at criticality for any of the families that we have considered 
(that is, $M^\star$ includes only the mass associated with that
portion of the overall matter distribution which appears to be static at 
criticality).
Moreover, it is more natural and convenient to use central proper time, $\tau$, rather 
than $t$ itself in the analysis.  Thus, the results below are described using
rescaled coordinates, $\tilde{\tau}$ and $\tilde{r}$, defined by
\begin{eqnarray}
\tilde{\tau} = \frac{\tau}{M^\star} \, ,\label{eq:deftautilde}\\
\tilde{r} = \frac{r}{M^\star} \, .\label{eq:defrtilde}
\label{eq:deftildecoord}
\end{eqnarray}
We note that under 
the scaling~(\ref{eq:scaletransformationt})--(\ref{eq:scaletransformationr}) the angular momentum transforms 
as 
\be
l \rightarrow kl \, .
\label{eq:lscale}
\ee 


The process we use to generate near-critical solutions is completely standard 
for this type of work.  All of the family definitions described above and 
summarized in Table~\ref{table:generic} contain multiple parameters that can be used 
to tune to the black hole threshold and, consistent with what has been 
found in many other previous studies of black hole critical phenomena, we find 
that which particular parameter is actually varied is essentially irrelevant
for the results.  Having chosen {\em some} specific parameter, $p$, to 
vary, any critical search begins by determining an initial 
bracketing interval, $[p_l, p_h]$, in parameter space such that evolutions 
with $p_l$ and $p_h$ lead to dispersal and black hole formation, respectively.
We then narrow the bracketing interval using a bisection search on $p$, predicating 
the update of $p_l$ or $p_h$ on whether or not a black hole forms.  The search 
is continued until $(p_h - p_l)/p_h \sim 10^{-15}$, so that $p^\star$ is computed 
to about machine precision (8-byte floating point arithmetic).  The value of 
$p_l$ at the end of this process corresponds to what we dub the marginally 
sub-critical solution.

\begin{figure}
\begin{center}
\includegraphics[width=0.45\textwidth]{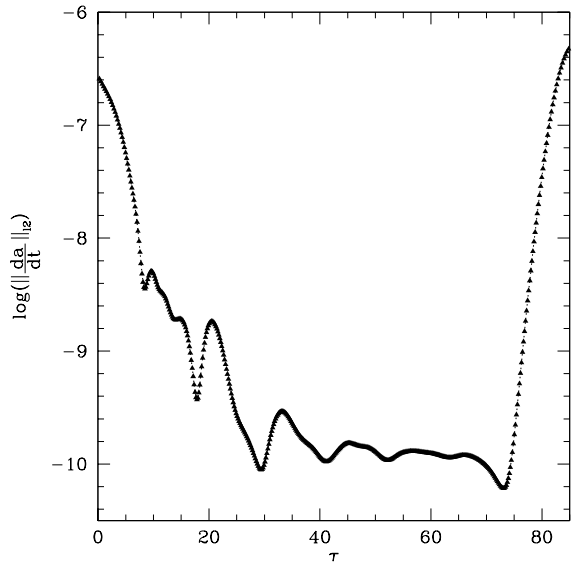}
\caption[Staticity of Critical Solution]{  \label{fig:adot}  
Time evolution of $\Vert \partial_t a(t,r) \Vert_2$ from a marginally sub-critical
calculation using family G1.  The plot provides strong evidence that the 
geometry of the threshold solution is static, a characteristic feature of type I 
behaviour.
}
\end{center}
\end{figure}

Quite generically, 
as we tune any family to a critical value $p^\star$, the 
phase space distribution function 
appears to settle down to a static solution which, as $p\to p^\star$, persists 
for a time that is long compared to the characteristic timescale for implosion and 
subsequent dispersal of the particles in the weakly-gravitating limit.
Representative illustrations of this behaviour are shown for marginally
sub-critical evolutions from two distinct initial 
data families in Fig.~\ref{fig:dynamic_phasespace} (family G8 in Table~\ref{table:generic}) and 
Fig.~\ref{fig:dynamic_phasespace2} (family G10).
Similarly, the 
spacetime geometry--encapsulated in the metric functions $a$ and $\alpha$---also 
becomes increasingly time-independent as criticality is approached. 
Fig.~\ref{fig:adot} displays the evolution of the $\ell_2$-norm of the time 
derivative of $a$ during marginally sub-critical evolution for family G1. 
We thus have strong evidence that the critical solutions that we are  finding
are static---characteristic of type I critical behaviour---and consistent with what 
has been observed previously for the case of the {\em massive} Einstein-Vlasov system.
\begin{figure}
\begin{center}
\includegraphics[width=0.45\textwidth]{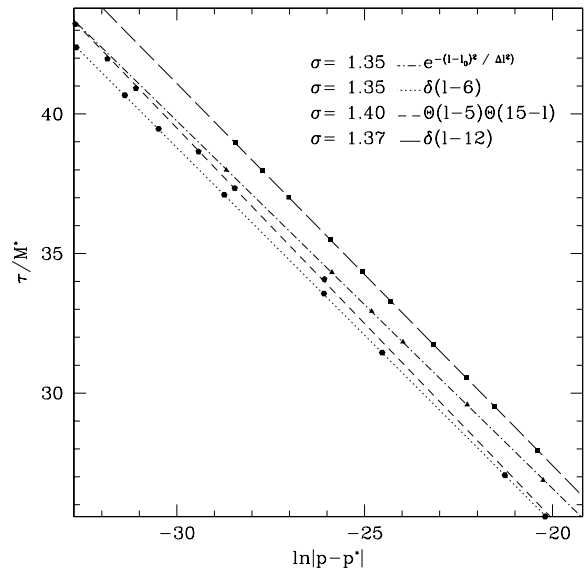}
\caption[Time Scaling]{\label{fig:timescale}  
Lifetime scaling of near-critical configurations for families G8, G1, G10 and G4 (top
to bottom and noting that G10 and G8 are 3D calculations while the others are 2D).  Here the symbols 
plot estimates of the amount of time the state of the 
system is well approximated by the static critical solution---measured in units
of the rescaled proper time defined by~(\ref{eq:deftautilde})---as a function of $\ln|p-p^\star|$.
The lines are least squares fits to 
$\tau = -\sigma \ln|p-p^\star|$ where $\sigma$ is the reciprocal of the eigenvalue
(Lyapunov exponent) corresponding to the presumed single growing mode of 
the critical solution.  To the estimated level of accuracy in our calculations the 
measured values of $\sigma$ are the same for the three families. However, we cannot 
state with certainty that there is precise universality in this regard.
}
\end{center}
\end{figure}
\begin{figure}
\begin{center}
\includegraphics[width=0.45\textwidth]{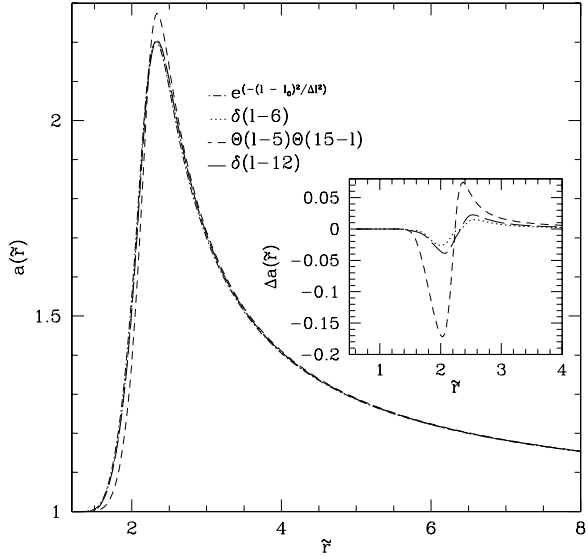}
\caption[Critical Geometry]{  \label{fig:final_geometry}  
Radial metric function $a(\tilde{r})$ at criticality for families G8, G1, G10 
and G4.  The results plotted here, together with those displayed in 
Fig.~\ref{fig:final_geometry2}, show that there is relatively little variation in the geometry of 
the static critical configuration as a function of the specifics of the initial
data.  The inset plots the deviation in $a$ for families G1, G10 and G4 relative 
to G8. 
}
\end{center}
\end{figure}
\begin{figure}
\begin{center}
\includegraphics[width=0.45\textwidth]{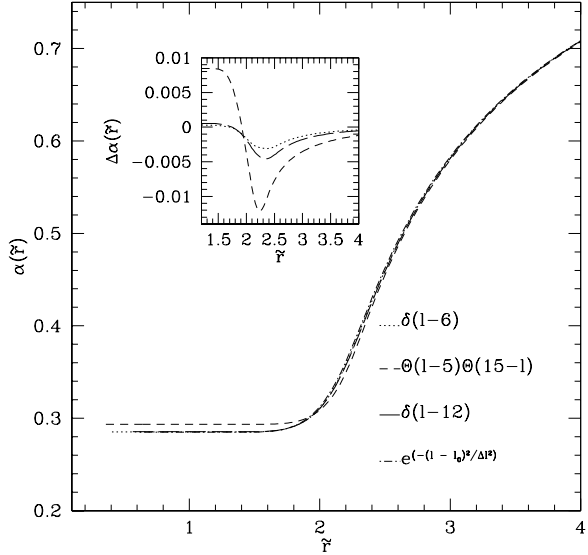}
\caption[Critical Geometry]{  \label{fig:final_geometry2}  
Lapse function $\alpha(\tilde{r})$ at criticality for families G8, G1, G10 
and G4.  The comments made in the caption of the previous figure apply 
here as well.
}
\end{center}
\end{figure}


Further evidence for generic type I transitions in the model is provided by observations of 
lifetime scaling 
of the form~(\ref{eq:deftimescaling}) near criticality, which is expected if the 
critical solutions are one-mode unstable.  Typical results from calculations using
%
%
families G1, G4, G8 and G10 are shown in Fig.~\ref{fig:timescale}:
the linearity of the lifetime of the static critical configuration as 
a function of $\ln|p-p^\star|$ is apparent.  We have observed such scaling for all of the 
families that we have studied (in both the 2D and 3D cases) and 
Table~\ref{table:timescaling} provides a summary of the  measured values of the scaling 
exponent, $\sigma$.

We note that the specific form of the matter configuration at  criticality 
exhibits significant dependence on the family of initial data
that is used to generate the critical solution.
This can be seen, for example, by comparing the 
last frames of Figs.~\ref{fig:dynamic_phasespace} and~\ref{fig:dynamic_phasespace2}.
On the other hand, as illustrated in Fig.~\ref{fig:final_geometry} and Fig.~\ref{fig:final_geometry2},
the {\em geometry} of the critical state is relatively insensitive to 
the initial conditions.

The spacetime geometry can be characterized by the central red shift, $Z_c$ defined by~(\ref{eq:defredshift}), and 
the unitless compactness parameter,
$\Gamma$, defined by
\be
\Gamma = {\rm max}_r\frac{2m}{r} \, .
\label{eq:defgam}
\ee
For the families considered in this section the values of $\Gamma$ and $Z_c$ fall in the ranges
\be
0.79 \lesssim \Gamma \lesssim 0.81 \, ,
\label{eq:rangeG}
\ee
\be
2.4 \lesssim Z_c \lesssim 2.5 \, .
\label{eq:rangeZ}
\ee
As discussed in the next section, 
these ranges are relatively small in comparison to those found in our investigation of critical
behaviour using nearly-static initial data.

What is striking about the results assembled in Table~\ref{table:timescaling} is that there appears to be
a small variation, at most, in the time scaling exponent associated with the critical solutions 
produced from our generic initial conditions.  Specifically, the data is consistent with
\be
\sigma = 1.4 \pm 0.1 \, ,
\label{eq:finalsigval}
\ee
and we emphasize that this concordance arises 
despite the significant observed variation in the phase-space distribution of the particles
among the various critical solutions.
%
\begin{table}
\begin{center}
{\footnotesize
\begin{tabular}{| c | c | c | c | c | c |}
 \hline
 Family & $l_0$ & $\sigma$ & Family & $l_0$  & $\sigma$  \\ \hline
 G1     & 5     &  $1.32 \pm 0.08$ & G3     & 12    &  $1.36 \pm 0.06$\\ \hline
 G1     & 6     &  $1.35 \pm 0.07$ & G4     & 12    &  $1.37 \pm 0.05$\\ \hline
 G1     & 7     &  $1.36 \pm 0.06$ & G5     & 12    &  $1.44 \pm 0.06$\\ \hline
 G1     & 8     &  $1.33 \pm 0.06$ & G6     & 12    &  $1.43 \pm 0.04$\\ \hline
 G1     & 9     &  $1.33 \pm 0.06$ & G7& $6$ \& $12$&  $1.37 \pm 0.07$\\ \hline
 G1     & 10    &  $1.32 \pm 0.06$ & G8     & 10    &  $1.35 \pm 0.05$ \\ \hline
 G1     & 11    &  $1.35 \pm 0.05$ & G9     & 10    &  $1.36 \pm 0.05$\\ \hline
 G1     & 12    &  $1.37 \pm 0.05$ & G10    & -     &  $1.40 \pm 0.05$\\ \hline
 G2     & -     &  $1.36 \pm 0.07$ &        &       &                 \\ \hline
\end{tabular}
}
\caption{
Summary of measured lifetime scaling exponents for the massless Einstein-Vlasov 
model from experiments using the various initial data families enumerated in Table~\ref{table:generic}.
In addition to the overall functional form of the initial distribution functions, a key 
parameter that varies among the sets of calculations is $l_0$, which 
is the angular momentum of any and all particles for families G1, G2--G6 (2D) and the 
center of the angular momentum distribution for families G8 and G9 (3D).  ($l_0$ is the 
tuning parameter for G6, and family G7 is another 
special case where the initial data is comprised of a superposition of two shells of particles,
each having a distinct angular momentum parameter.  Since angular momentum is a conserved 
quantity there is no mixing of the two distributions during the evolution.)
For simplicity
of presentation we have not listed the other parameters defining the different initial
configurations.
Quoted uncertainties in the values of $\sigma$ are based on 
variations in the total mass of the system during the evolutions and comparison with
results computed at lower resolution.
Typical grid sizes used for the listed results 
are $n_r \times n_p = 1024 \times 1024$ (2D) or $n_r \times n_p \times n_l = 256 \times 128 \times 64$
(3D).
To the level of accuracy in our calculations we find consistency with 
a single value of the scaling exponent, $\sigma = 1.4 \pm 0.1$.
}
\label{table:timescaling}
\end{center}
\end{table}
\subsection{Near-static massless case\label{sec:staticmassless}}

Our 
second approach to study critical solutions in the massless Einstein-Vlasov system starts with
the construction of static initial data using the procedure described in Sec.~\ref{sec:initstat}.
We specialize the general form~(\ref{eq:staticconf}) to 
\be
\Phi(E,l)= C (1-E/E_0)^{b}\Theta(E_0-E)\delta(l-l_0) \, ,
\label{eq:phi_form} \ee
where $E_0$ is a given cutoff energy and $C$, $b$ and $l_0$ are additional adjustable parameters.
Here we focus exclusively on the case of fixed angular momentum (2D calculations) since 
the results of the previous section
suggest that the essential features of the critical solutions are not 
significantly dependent on whether or not $f$ has non-trivial dependence on  $l$.
%
In addition, from the scale free symmetry in the system (see (\ref{eq:scaletransformationt})~and~(\ref{eq:lscale})),
we can conclude that
varying the value of angular momentum is equivalent to rescaling the radial coordinate. Therefore,
without loss of generality we can set $l$ to an arbitrary fixed value, eliminating one 
of the parameter-space dimensions in our surveys.
Additionally, so that we can meaningfully compare results from different initial conditions,
we again rescale the radial coordinate 
by the total mass of the system~(\ref{eq:deftildecoord}).
%
Furthermore, by virtue of the transformation~(\ref{eq:escaling}), the static 
profiles depend on $E_0$ 
only through the ratio $E_0/\alpha_0$ and,
since it simplifies the numerical 
analysis, we actually use this ratio as one of the control parameters.


For specified values of the free parameters $C$, $b$ and $E_0/\alpha_0$,
we integrate equations (\ref{eq:a_func})--(\ref{eq:Trr_func}) outward 
until we reach a radial location, $r_X$, where the
particle density $\Phi(E,l)$ vanishes.  We then extend the solution for $a$ and $\alpha$
to the outer boundary of the computational domain by attaching a Schwarzschild geometry with 
the appropriate mass.

%
%
%
We note that not all choices of the three free parameters lead to distribution functions with
compact support---that is, with $f(0,r,p_r)\equiv 0$ for $r$ greater than some 
$r_X$---so that the configuration
represents a single shell of particles.
Indeed, by examining the expression for the particle energy in the massless case:
\be
E(r,p_r) = \alpha(r) \sqrt{(p_r/a)^2 + (l/r)^2} \, ,
\ee 
we see that, for $p_r$ sufficiently small, $E(r,p_r)$ can remain below the cutoff $E_0$ for 
large $r$.  In practice 
this will 
yield solutions with multiple shells, where $\Phi$ 
vanishes at $r_X$, but then becomes non-zero on a infinite number of intervals 
in $r$ (in general these intervals can be disjoint or contiguous, as has previously
been seen in \cite{And_Rein:2007}).  Although it might be 
interesting to consider the critical dynamics of multiple-shell solutions, we do not 
do so here.  
 We also note that for given values of $b$ and $E_0/\alpha_0$ we find solutions with a distinct shell (i.e.~where
$\Phi$ {\em does} vanish at some radius) only for a certain range of $C$, but that range can span several
orders of magnitude.

\begin{figure}
\begin{center}
\includegraphics[width=0.23\textwidth]{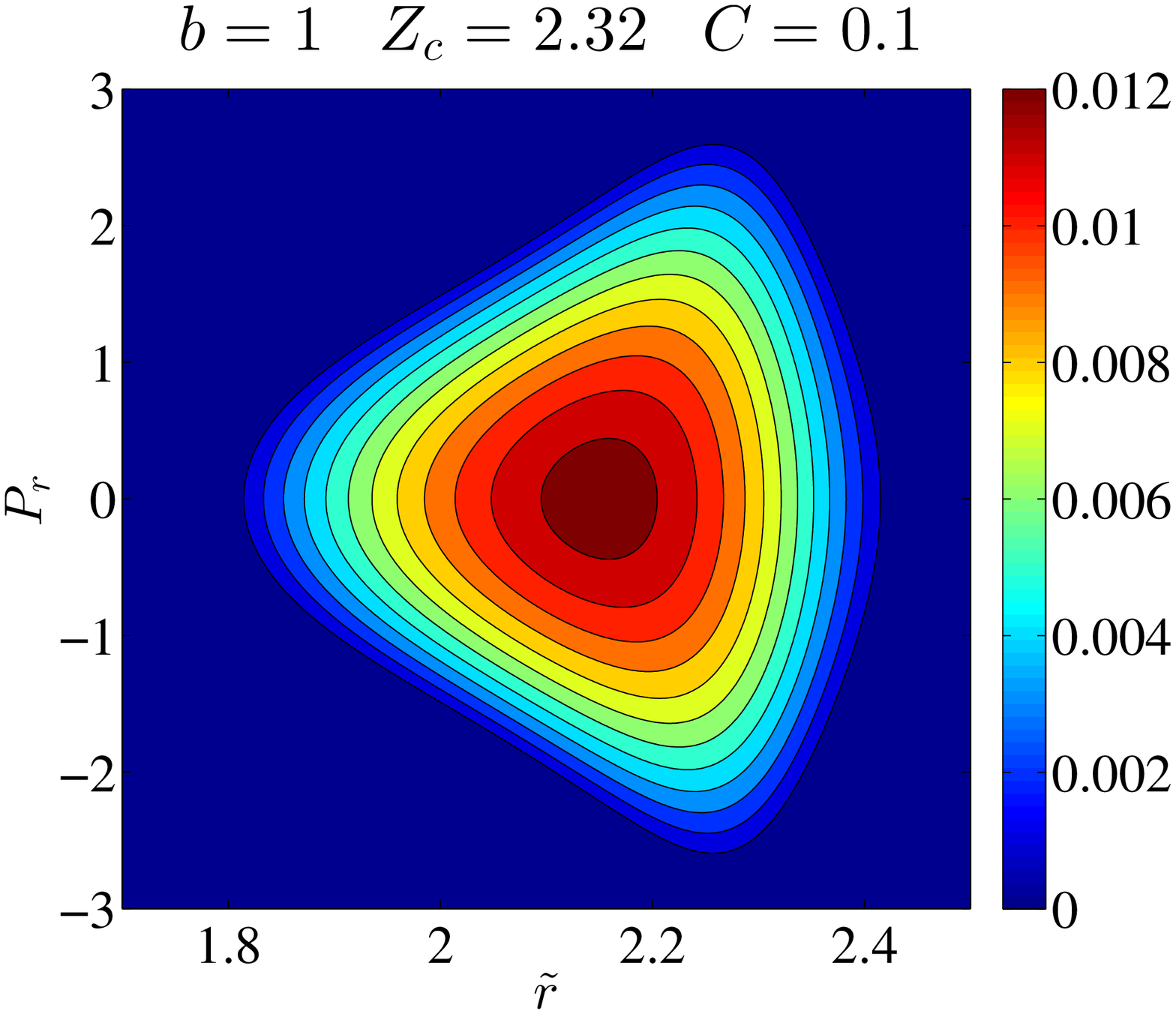}
~\includegraphics[width=0.23\textwidth]{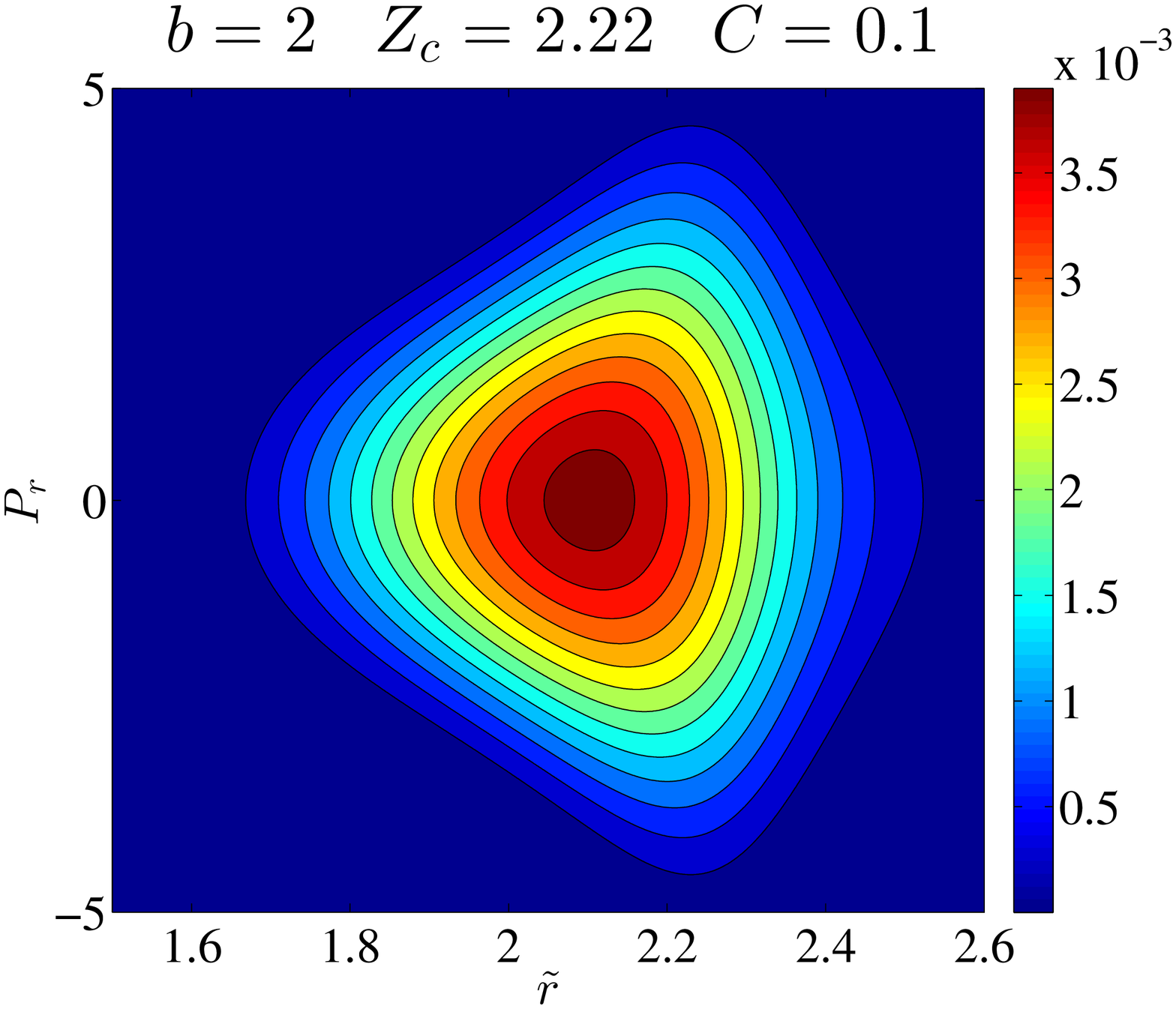}\\
\includegraphics[width=0.23\textwidth]{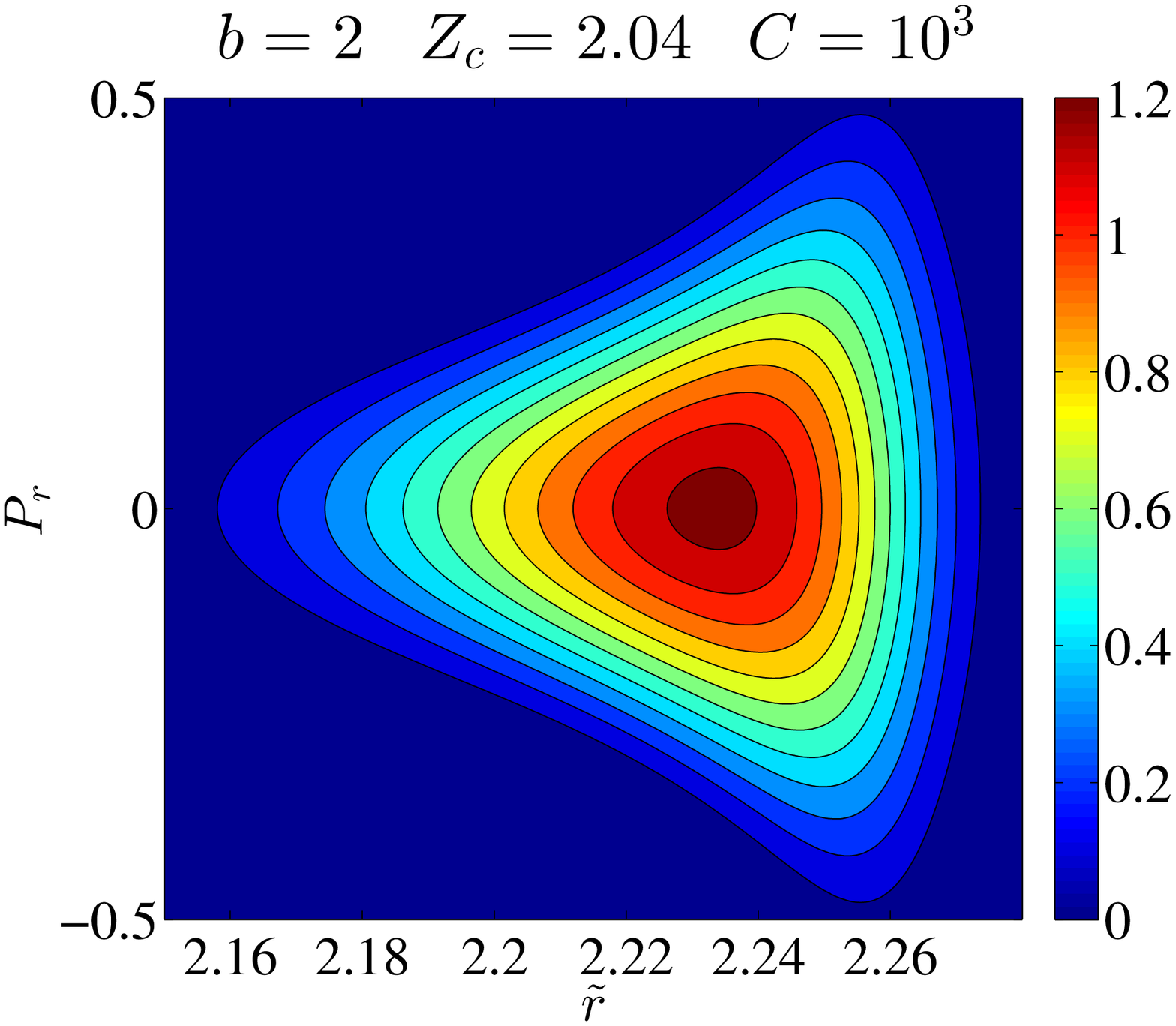}~
\includegraphics[width=0.23\textwidth]{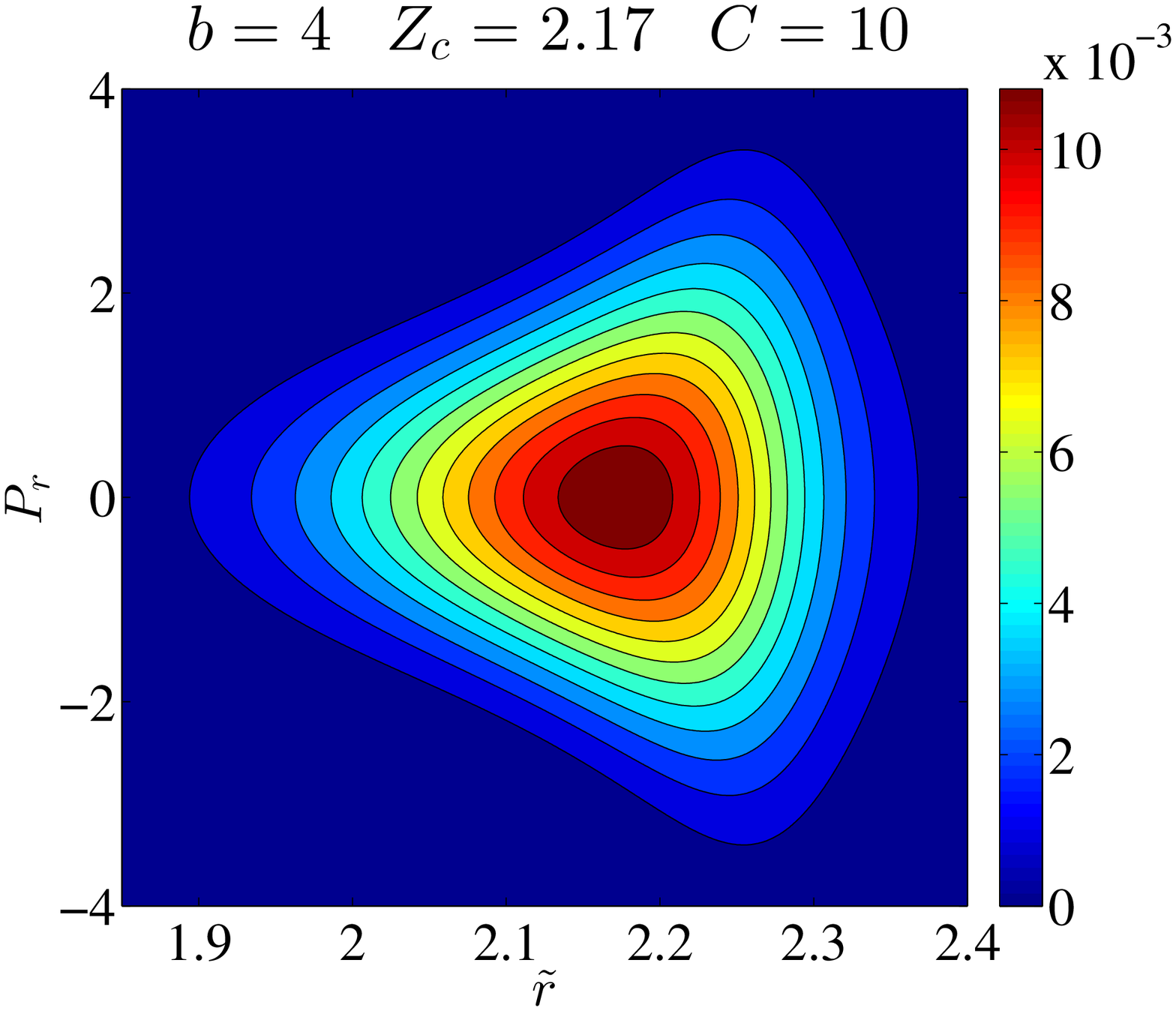}
\caption[Phase Space Configurations]{  \label{fig:static_phasespace}  
 Sample 
 static phase space configurations computed from the ansatz~(\ref{eq:phi_form})
 using different choices of adjustable parameters. Note that although we use the rescaled 
 radial coordinate ${\tilde r}$ in all of the plots, the ranges in ${\tilde r}$, $p_r$ and 
 $f$ vary from frame to frame.  Clearly, there is a strong dependence 
 of $f$ on the chosen parameter values.  As described in more detail in the 
 text, for any given values of $b$ and $Z_c$ there 
 is a finite range of $C$  for which we find static solutions where $f$ has compact support.
}
\end{center}
\end{figure}

\begin{figure}
\begin{center}
\includegraphics[width=0.45\textwidth]{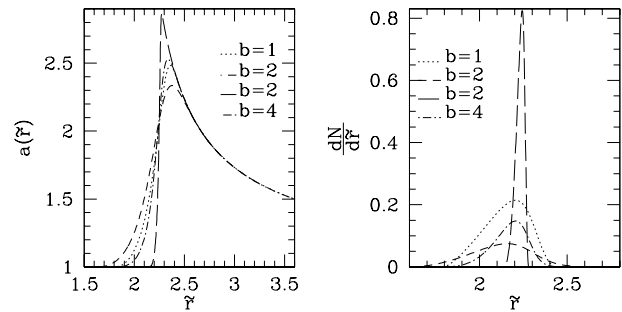}
\caption[Static Geometries]{\label{fig:static_geometry}  
Plots of the radial metric function, $a(r)$, and differential particle number,
$dN(r)/dr$, for the configurations shown in Fig.~\ref{fig:static_phasespace}.
The graphs of $dN(r)/dr$ highlight the fact that the critical solutions are shell-like, with a thicknesses
and effective densities that are strongly dependent on the choice of parameters in~(\ref{eq:phi_form}).
}
\end{center}
\end{figure}

Fig.~\ref{fig:static_phasespace} shows the distribution function for four sample static 
configurations constructed as described above, with the associated geometrical variables plotted 
in Fig.~\ref{fig:static_geometry}.   
Relative to the apparently static solutions generated by tuning generic initial data, 
the family-dependence of both the distribution function and metric variables here
is much more pronounced.

\begin{figure}
\begin{center}
\includegraphics[width=0.45\textwidth]{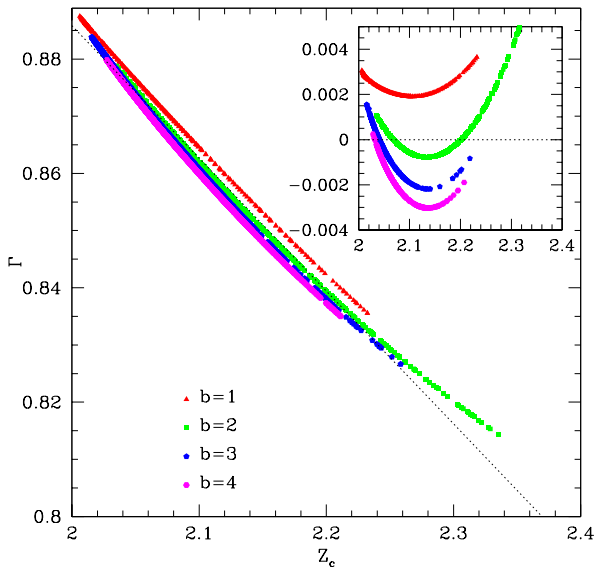}
\caption[Static Family of Solutions]{  \label{fig:static_family}  
 The value of $\Gamma={\rm max}_r(2m/r)$ versus central redshift, $Z_c$, for various
static solutions.  Each set of points comprises several thousand distinct solutions
and comes from a two-dimensional parameter
space survey, in which both $C$ and $E_0/\alpha_0$ are varied. Although for given $b$ and 
$E_0/\alpha_0$ we can only find 
acceptable static solutions in certain ranges of $C$, those ranges can span several 
orders of magnitude.  However, for fixed $b$ the solutions tend to collapse to near-linear loci in $Z_c$--$\Gamma$ space,
and the inset graph, which plots the deviation of the data from a linear least squares fit, is intended 
to emphasize this behaviour.  More detailed examination of the data suggests that the configurations do 
{\em not} lie precisely along one-dimensional curves, but additional study would be required to 
determine whether this is really the case.
The solutions apparently satisfy the Buchdahl
inequality $\Gamma < 8/9$ (also seen in the calculations reported in~\cite{And_Rein:2007} 
for the 
massive
case), 
as is expected from Andr\'{e}asson`s rigorous results~\cite{And:onBuch}.  Moreover,
there also seems to be a {\em lower} bound on the compactness,
$\Gamma \sim 0.81$.
}
\end{center}
\end{figure}

One interesting way of characterizing the static solutions is to plot the compactness parameter, $\Gamma$, defined
by (\ref{eq:defgam}), 
as a function of the central redshift, $Z_c$.   We do this for a large number of configurations in 
Fig.~\ref{fig:static_family} where, as described in more detail in the caption, each set of points results 
from a two-dimensional parameter space survey wherein both $E_0/\alpha_0$ and $C$ are varied.  
The fact that the solutions from each of these surveys tend to ``collapse'' to one-dimensional curves 
in $Z_c$--$\Gamma$ space is striking and we do not have any argument at this time for why this should 
be so.

All of the static solutions that we have found satisfy Buchdahl's inequality, $\Gamma < 8/9$, 
originally derived in
the context of fluid matter~\cite{Buchdahl}, and the most compact configurations are quite close to that
limit.  
Here it is crucial to note that Andr\'{e}asson
has proven rigorously that the Buchdahl inequality 
is satisfied by any static solution of the spherically symmetric Einstein-Vlasov system~\cite{And:onBuch}.
Further, he has demonstrated that one can construct static shell-like configurations which, in the 
limit of infinitesimal thickness in $r$, can have $\Gamma$ arbitrarily close to $8/9$.
Although  not explicitly mentioned 
in~\cite{And:onBuch}, it is clear that his proof is valid for $m=0$.  
Given the nature of 
Andr\'{e}asson`s result, the observation that our solutions satisfy the bound clearly amounts to little
more than additional evidence that our calculations are faithful to the model under study.
However it {\em is} interesting that the highest 
values of $\Gamma$ seen in Fig.~\ref{fig:static_family}---and which plausibly {\em are} approaching
$8/9$---are associated with very thin shell-like solutions.
Additionally, for the configurations we have studied (not all of which are represented in 
Fig.~\ref{fig:static_family}) there is apparently also a lower bound on the compactness,
$\Gamma \sim 0.81$.  Finally, the ranges of $\Gamma$ and $Z_c$ spanned by the explicitly 
static solutions 
\be
0.80 \lesssim \Gamma  \lesssim 0.89 \, ,
\ee
\be
2.0 \lesssim Z_c \lesssim 2.4 \, ,
\ee
are larger than those seen for the tuned generic data,
consistent with the comment above concerning the relatively large variations in the 
metric variables as well as the distribution function.


\begin{table}
\begin{center}
  \begin{tabular}{| c | c | c | c | c | c | }
    \hline
	 $b$ & $Z_c$  & $C$   & $\delta f$    & $\sigma$  \\ \hline
	 $1$ & $2.32$ & $0.1$ & $\delta f_1$  & $1.45 \pm 0.05$ \\ \hline
	 $1$ & $2.23$ & $0.3$ & $\delta f_1$  & $1.45 \pm 0.04$ \\ \hline
	 $2$ & $2.22$ & $0.1$ & $\delta f_1$  & $1.43 \pm 0.04$ \\ \hline
	 $4$ & $2.17$ & $10$ & $\delta f_1$  & $1.43 \pm 0.04$ \\ \hline
	 $2$ & $2.35$ & $0.1$ & $\delta f_1$  & $1.40 \pm 0.05$ \\ \hline
	 $2$ & $2.35$ & $0.1$ & $\delta f_2$  & $1.40 \pm 0.05$ \\ \hline
	 $2$ & $2.35$ & $0.1$ & $\delta f_3$  & $1.40 \pm 0.05$ \\ \hline
  \end{tabular}
\caption[Time Scaling Static Case]{Measured lifetime scaling exponent for explicitly 
static solutions constructed from ansatz~(\ref{eq:phi_form}) with
various choices of the adjustable parameters $b$, $E_0/\alpha_0$ and $C$ ($Z_c$ is effectively 
controlled by $E_0/\alpha_0$, but is determined {\em a posteriori}), and the different 
types of perturbations, $\delta f$, enumerated in~(\ref{eq:df1})--(\ref{eq:df3}).
Proceeding from the assumption that the static solutions {\em are} characterized by 
a single unstable mode, we anticipate that the computed value of $\sigma$ associated 
with a specific configuration (i.e.~for given $b$, $Z_c$ and $C$) should be independent
of the form of $\delta f$, and this is precisely what we observe (compare rows 1 and 2, 
and 5, 6 and 7).
However, we also see once again that there is little, if any, variation in the scaling 
exponent with respect to the underlying critical solution:  the results in the table are 
consistent with 
$\sigma =1.43 \pm 0.07$ 
}
\label{tb:timescale_static}
\end{center}
\end{table}

Using our evolution code, 
we investigate the relation of the explicitly-static solutions to critical 
behaviour in the model as follows.  For initial conditions we
set 
\be
f(0,r,p_r,l^2) = f^0(r,p_r,l^2) + (A-1)\delta f(r,p_r,l^2)  \, ,
\ee
where $f^0$ is a static configuration, $\delta f(r,p_r,l^2)$ is some given perturbation 
function with 
at least roughly the same support as $f^0$, and $A$ is a tunable parameter which controls
the amplitude of the perturbation. Clearly, $A=1$ results in initialization with 
the static solution itself.  We have experimented with the following three 
choices for the perturbation function:
\be
\delta f_1(r,p_r,l^2) = f^0(r,p_r,l^2) \, ,
\label{eq:df1}
\ee
\be
\delta f_2(r,p_r,l^2) = \sin\left(\frac{2\pi f^0(r,p_r,l^2)}{f_{\rm max}}\right) \, ,
\label{eq:df2}
\ee
\be
\delta f_3(r,p_r,l^2) = f^0(r,p_r,l^2)(f_{\rm max}-f^0(r,p_r,l^2))p_r \, ,
\label{eq:df3}
\ee
where
$f_{\rm max}$ is the maximum of $f^0$ over the computational domain. We then 
perform standard tuning experiments in which we vary $A$ to isolate a 
threshold solution.

\begin{figure}
\begin{center}
\includegraphics[width=0.45\textwidth]{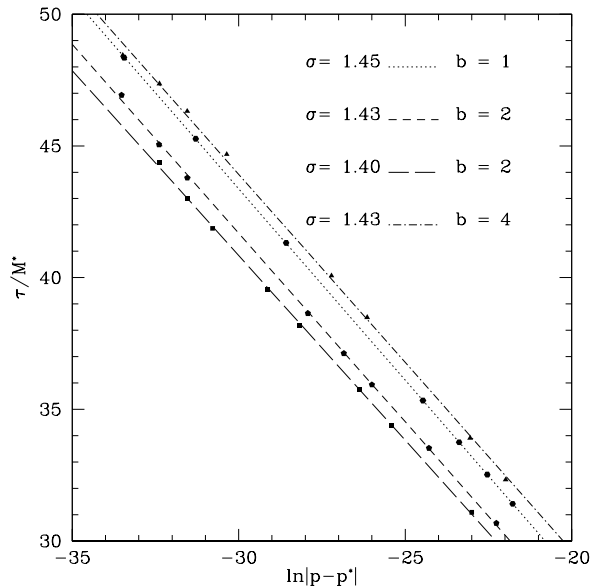}
\caption[Static Family Time Scaling]{\label{fig:static_timescaling}
Lifetime scaling computed from families of initial data based on the static
configurations plotted in 
Figs.~\ref{fig:static_phasespace} and \ref{fig:static_geometry}.
The tuning parameter in this instance controls the amplitude of a perturbation
that is added to the base solution (here we used the form $\delta f_1$~(\ref{eq:df1}))
and, in all cases, the sign of the perturbation determines 
whether the evolution leads to dispersal or black hole formation.  The results shown
here provide evidence that the static configurations calculated from the ansatz~(\ref{eq:phi_form})
act as type I critical solutions.  Additionally, we see that there is very little 
variation in the measured scaling exponents, $\sigma$, which are again determined via least
squares fits to~(\ref{eq:deftimescaling}).
}
\end{center}
\end{figure}

Interestingly, we find strong evidence that {\em all} of the static solutions 
based on~(\ref{eq:phi_form}) 
that we have found sit at the threshold of 
black hole formation, so that setting
$A>1$ results in black hole formation while taking $A<1$ results in complete dispersal of the 
matter (or vice versa, dependent on the precise form of $\delta f$).
As should be suspected then, and as is shown for four families in Fig.~\ref{fig:static_timescaling},
the solutions generated by 
dynamically evolving the perturbed static configurations exhibit time scaling---this strongly suggests that 
the time-independent
solutions are all one-mode unstable.
Table~\ref{tb:timescale_static} provides a summary of the time-scaling exponents we have measured 
for a set of experiments based on four distinct static solutions and the three different types of 
perturbation defined by~(\ref{eq:df1})--(\ref{eq:df3}).

As was the case for the generic families, the measurements here indicate that although the static 
solutions display significant variation in both the distribution function and geometric variables,
there is little variation in the scaling exponent.  Here we find
\be
\sigma = 1.43 \pm 0.07 \, .
\ee
Recalling~(\ref{eq:finalsigval}), and given the estimated uncertainty in our calculations, we can
not exclude the possibility that $\sigma$ is truly universal for 
the massless-sector critical solutions
which we have constructed.
Particularly given the variation in the spacetime geometries 
involved, constancy of the eigenvalue of the unstable mode associated with criticality
would be truly remarkable.  However, even if $\sigma$ does span some finite range, the 
apparent tightness of that range is an aspect of critical behaviour in the massless system
that begs understanding.

Finally, we note that the static critical solutions from the generic calculations are 
characterized by compactness, $\Gamma \sim 0.8$,  which is at the low end of the range spanned by
the explicitly static solutions.  We do not yet know whether a more extensive parameter 
space survey of generic data could produce critical configurations with larger $\Gamma$, and it 
would be interesting to further investigate this issue.



\subsection{Generic massive case \label{sec:gen_massive}}

Following previous studies~\cite{Rein-Randall:1998,Inaki:2001,stevenson}, we have also examined the 
case where the particles have rest mass and find results that are in general agreement with
the earlier work, including strong evidence for the existence of static solutions at 
the black hole threshold that 
exhibit lifetime scaling.
However, we note that in both  \cite{Inaki:2001} and \cite{stevenson} the 
initial data configurations were kinetic energy dominated. 
For example, a typical calculation in~\cite{Inaki:2001} used unit particle 
mass and $f(0,r,p_r,l)$  which was gaussian in the three coordinates with characteristic 
values $r \sim 3$, $p_r \sim 1$ and $l\sim3$.
From expression~(\ref{eq:energy}) for the particle energy we can thus infer that
the initial data sets had kinetic energy about an order
of magnitude larger than rest mass energy.  Thus we expect that 
those previous results should be similar to what 
we see for massless particles.
Indeed, taking into account the different time parameterization used ($t$ normalized to coincide 
with property time at infinity), 
the scaling exponents quoted in~\cite{Inaki:2001} are consistent with our results.

Table~\ref{table:timescaling_massive} lists the values of the time scaling exponent we 
have determined in the massive case for the various types of initial data defined in Table~\ref{table:generic}.
We note that the initial data families that are used include ones that are very similar
to those adopted in~\cite{Inaki:2001} and \cite{stevenson}.
We see that the time scaling exponents are in fact close to those measured in the massless calculations, 
although the spread in the values is noticeably larger here 
(as it was in~\cite{Inaki:2001} and \cite{stevenson}). 
This increased spread is almost certainly 
due to the particle mass---i.e. the evolutions are not {\em completely} kinetic energy 
dominated.
%

Paralleling what was done in~Sec.\ref{sec:staticmassless}, as well as in~\cite{And_Rein:2007},
we can use perturbations of our
explicitly static solutions in the massive sector
to investigate critical behaviour.  Here 
there is a larger function space of static configurations, especially
since we can construct solutions with positive binding energy, $E_b$, defined by
\be
E_b \equiv M_0 - M \, ,
\ee 
where $M_0$ is the total rest mass and $M$ is the ADM mass.  Moreover, we can
build parameterized sequences of solutions that transition between positive and 
negative $E_b$, completely analogously to what can be done for perfect fluid 
models of general relativistic stars.
As in the perfect fluid case, we anticipate that: 1) solutions with $E_b > 0$ will be perturbatively 
stable, 2) there will be a change of stability at $E_b = 0$, and 3) for at least some 
range of $E_b < 0$, the static configurations will be one-mode unstable, and thus should 
constitute type I critical solutions.
We have performed additional calculations that confirm these expectations.
In particular, we were able to build a static solution
with $E_b$ negative, but relatively close to 0, which {\em did} lie at the black hole threshold and 
which had an associated scaling exponent $\sigma = 3.0 \pm 0.1$.  This value of $\sigma$ is clearly 
distinct from those listed in Table~\ref{table:timescaling_massive}. Thus, in contrast to the 
massless case where we can not conclusively state anything about possible variations in $\sigma$ for
type I critical solutions, we are confident that $\sigma$ {\em is} is not universal in the massive
case.  
In fact, were we able to construct static configurations with $E_b \rightarrow 0^{-}$, we 
assume that we would find $\sigma \rightarrow \infty$.  
Again, these observations and 
conjectures are entirely consistent with previous studies of 
the Einstein-Vlasov system, as well as work with gravitationally compact stars modelled with 
perfect fluids or bosonic matter.

\begin{table}
\begin{center}
{\footnotesize
\begin{tabular}{| c | c | c | c || c | c | c | c |}
 \hline
 Family & $l_0$ &$Z_c$     &  $\sigma$        & Family & $l_0$ &$Z_c$ & $\sigma$ \\ \hline
 G1     & 5     &2.47      &  $1.32 \pm 0.14$ & G1     & 12    &2.28  &  $1.46 \pm 0.07$\\ \hline
 G1     & 6     &2.39      &  $1.47 \pm 0.13$ & G2     & -     &2.39  &  $1.44 \pm 0.09$  \\ \hline
 G1     & 7     &2.31      &  $1.44 \pm 0.08$ & G3     & 9     &2.29  &  $1.54 \pm 0.07$ \\ \hline
 G1     & 8     &2.37      &  $1.49 \pm 0.08$ & G4     & 9     &2.43  &  $1.49 \pm 0.08$\\ \hline
 G1     & 9     &2.41      &  $1.49 \pm 0.08$ & G8     & 10    &2.24  &  $1.38 \pm 0.14$\\ \hline
 G1     & 10    &2.34      &  $1.48 \pm 0.07$ & G9     & 10    &2.41  &  $1.59 \pm 0.15$\\ \hline
 G1     & 11    &2.23      &  $1.54 \pm 0.07$ &        &       &      &                 \\ \hline
  \end{tabular}
}
  \caption{
Summary of measured lifetime scaling exponents for the massive Einstein-Vlasov 
model from experiments using the various initial data families enumerated in Table~\ref{table:generic}.
The results quoted here derive from calculations that parallel those described in 
Table~\ref{table:timescaling} for the massless system.  In contrast to the massless case, the observed 
variation in $\sigma$ is significant.
\label{table:timescaling_massive}
}
\end{center}
\end{table}

\section{Summary and Discussion\label{conclusions}}
We have constructed a new numerical code to evolve the Einstein-Vlasov system in spherical symmetry 
using an algorithm where the distribution function $f(t,r,p_r,l^2)$ is directly
integrated using finite volume methods.  This approach eliminates 
the statistical uncertainty inherent in the particle-based techniques that have been used 
in previous studies.
To reduce computational demands at a given discretization or, more importantly,
to allow for higher resolution, we can also run the code in a 2D mode where $l^2$ is some 
fixed scalar constant so that $f$ depends on only $r$ and $p_r$.

We have used the code to perform extensive and detailed surveys of the critical behaviour in the 
model with a particular focus on the case where the particles are massless. We note that we 
are unaware of any previous dynamical numerical calculations pertaining to the massless sector.

Our results derive from two classes of initial configurations. In the first the initial states
represents imploding shells of particles well removed from the origin, while the second involves perturbations
of configurations that are precisely static by construction.  Although time-independent solutions of the 
massive system have been constructed and analyzed previously, to our knowledge the static states we have 
found in the massless sector are the first of their kind.
Within each class we have studied numerous 
specific forms for the initial data and, for the near-static calculations, the perturbations that are 
applied to generate the threshold behaviour.  In all cases we 
find strong evidence for a Type I critical transition including: 1) a finite black hole mass at 
threshold and 2) lifetime scaling of the form~(\ref{eq:deftimescaling}).  The observations 
are all consistent with the standard picture for Type I behaviour, namely a static critical
solution with one unstable perturbative mode. 
Here we emphasize that---as is the case for any 
numerical study of critical behaviour---it is very difficult to preclude the existence of additional
unstable modes. However, the degree to which the scaling laws
are satisfied suggests that if such modes do exist they have growth rates significantly 
smaller than the dominant one.


For generic initial data with massless particles,
we have found that there is a considerable variation in the morphology of $f$ among the different critical 
solutions we have computed and, to a lesser extent, in the details of the spacetime geometries encoded in $a(t,r)$ and $\alpha(t,r)$.
Interestingly though, there is relatively little variation in the time scaling exponents that we 
have measured: all seem to be in the range $\sigma = 1.4 \pm 0.1$.   

In the case of near-static initial conditions with $m=0$ the key results are quite 
similar.  Again, there is a large variation in the functional form of the distribution function at 
threshold. In this instance this can be seen as a direct reflection of the freedom inherent in 
the ansatz~(\ref{eq:phi_form}) 
which involves the specification of two essentially arbitrary functions.  
Not surprisingly, there is thus a more noticeable range in the geometries at criticality relative
to the generic calculations,
as can be clearly seen, for example, through
examination of quantities such as the compactness and central redshift.  Once again, however, we 
observe only a small dispersion in the measured scaling exponents. Specifically, across all 
near-static families that we have examined we find $\sigma = 1.43 \pm 0.07$.

Thus, considering {\em all} of the calculations that we have performed, we have indications of at least 
a weak form of universality of the time-scaling exponent in the massless Einstein-Vlasov model.  Here 
we note that as mentioned in the introduction, the calculations reported in~\cite{Inaki:2001} were also suggestive
of a universal value of $\sigma$ and perhaps of the critical geometry.  Those computations used a non-zero
mass and, as also discussed previously, the work of~\cite{And_Rein:2006,And_Rein:2007} established that 
the spacetime structure at criticality could {\em not} be universal in the massive model.  However, as noted in Sec.~\ref{sec:gen_massive} 
the initial data families used in~\cite{Inaki:2001} were kinetic energy dominated 
(effectively massless), and 
so there is no contradiction between what was seen there (and here) and~\cite{And_Rein:2006,And_Rein:2007}.

In all of our calculations, and in accord with Andr\'{e}asson's proof of the Buchdahl inequality
in the model~\cite{And:onBuch}, we observe that the gravitational compactness satisfies $\Gamma < 8/9$ , with 
thin shell-like solutions coming closest to saturating the bound.

We also want to emphasize an additional feature of the massless model that is 
apparent from our calculations: the particle angular momentum does not have a significant impact on
the features of the critical solution (apart from the obvious fact that the particles {\em do} have
angular momentum in all of our computations).
Heuristically, this can be at least partly ascribed  to 
the scaling symmetry~(\ref{eq:scaletransformationt})--(\ref{eq:scaletransformationr}).
The symmetry effectively reduces the number of free parameters---relative to a naive analysis---available
for variation in the search for critical solutions.
Specifically,
given
any distribution of the form $f(r,p_r)\delta(l-l_1)$, where $l_1$ is fixed,  
we can map to a distribution
$f'(r,p_r)\delta(l-l_2)$, with $l_1\ne\l_2$, which has an associated 
geometry that is diffeomorphic to the original.

%

Given that there is clearly {\em no} 
universality of the fundamental dynamical variables at threshold, the fact that the 
variation in $\sigma$ is, at most, small is
a feature of the calculations for which we currently have no explanation.
Additionally, as discussed in the introduction, the argument advanced in~\cite{PhysRevD.65.084026} 
suggests that there should be {\em no} type I behaviour in the Einstein-Vlasov system for either 
the massless or massive models.
At this time, we do not understand how---if at all---this argument can be reconciled with 
our current results and those from previous numerical studies.


A direct analysis of the perturbations of the critical solutions---especially
the precisely static ones---would be very helpful at this point.  Starting with the perfect-fluid work of 
Koike {\em et al}~\cite{Koike:1995jm}, perturbation analyses of 
the critical configurations in many different models have been extremely effective in advancing
our understanding of black-hole critical phenomena.
In particular, relative to measurements made through direct 
solution of PDEs and tuning experiments, perturbative methods can provide highly accurate values 
for the eigenvalues of the unstable modes (or, equivalently, for the scaling exponents).
However, in our case the task of explicitly constructing perturbations is significantly complicated by 
the fact that there is no one-to-one correspondence between the geometry and the phase-space distribution 
of the particles.  So far we have been unable to formulate a well-defined approach to computation
of the perturbations and will have to leave that for future work.  

Finally, it would be interesting to extend 
this work to the Einstein-Boltzmann system,
where the introduction of explicit interactions between particles
would provide the means to investigate the connection between criticality in 
phase-space-based models and hydrodynamical systems. This in turn might lead to a 
more fundamental understanding of critical collapse in fluid models.
\nocite{1989ApJ344146R}
\nocite{Gleiser:2009tx}

\begin{acknowledgments}
This research was supported by NSERC, CIFAR and by a Four Year Fellowship scholarship to 
Arman Akbarian from UBC. 
Calculations were performed using Compute Canada (Westgrid) facilities.
The authors thank William G. Unruh and Jeremy Heyl for 
insightful comments and discussions.
\end{acknowledgments}

\bibliography{paper}

\end{document}